\documentclass{ieeeaccess}
\usepackage{cite}
\usepackage{amsmath,amssymb,amsfonts}
\usepackage{algorithmic}
\usepackage{graphicx}
\usepackage{hyperref}
\hypersetup{
    colorlinks=true,
    linkcolor=blue,
    filecolor=magenta,      
    urlcolor=cyan,
}

\usepackage{textcomp}
\def\BibTeX{{\rm B\kern-.05em{\sc i\kern-.025em b}\kern-.08em
    T\kern-.1667em\lower.7ex\hbox{E}\kern-.125emX}}

\newcommand{\chg}[1]{{\color{red}{#1}}}

\newtheorem{theorem}{Theorem}[section]
\newtheorem{lemma}[theorem]{Lemma}

\newtheorem{definition}[theorem]{Definition}
\newtheorem{remark}[theorem]{Remark}

\newtheorem{assumption}[theorem]{Assumption}
\newcommand{\bld}[1]{\mathbf{#1}}
\DeclareMathOperator{\sign}{sign}

\begin{document}
\history{Date of publication xxxx 00, 0000, date of current version xxxx 00, 0000.}
\doi{10.1109/ACCESS.2017.DOI}

\title{Robust leader-following consensus of high-order multi-agent systems in prescribed time}
\author{
\uppercase{J. Armando Colunga}\authorrefmark{1},
\uppercase{H\'{e}ctor M. Becerra}\authorrefmark{1}, \IEEEmembership{Member, IEEE}, 
\uppercase{Carlos R. V\'azquez}\authorrefmark{2}, \IEEEmembership{Member, IEEE},
\uppercase{and David G\'omez-Guti\'errez}\authorrefmark{2,}\authorrefmark{3},
\IEEEmembership{Senior Member, IEEE}
}
\address[1]{Centro de Investigaci\'{o}n en Matem\'{a}ticas (CIMAT), Jalisco S/N, Col. Valenciana, 36023, Guanajuato, Mexico. (e-mails: jose.colunga@cimat.mx, hector.becerra@cimat.mx)}
\address[2]{Tecnol\'ogico de Monterrey, Escuela de Ingenier\'ia y Ciencias, Av. General Ram\'on Corona 2514, 45201, Zapopan, Jalisco, Mexico (e-mail: cr.vazquez@itesm.mx)}
\address[3]{Multi-agent Autonomous Systems Lab, Intel Labs, Intel Tecnología de México, Av. del Bosque 1001, 45019, Zapopan, Jalisco, Mexico (e-mail: david.gomez.g@ieee.org)}
\tfootnote{This work was supported in part by Intel Corporation. \\ 
\chg{This is the preprint version of the published manuscript: J. A. Colunga, H. M. Becerra, C. R. Vázquez and D. Gómez-Gutiérrez, ``Robust Leader-Following Consensus of High-Order Multi-Agent Systems in Prescribed Time," in IEEE Access, vol. 8, pp. 195170-195183, 2020, doi: 10.1109/ACCESS.2020.3033789.
Please cite the publisher’s version: \url{https://doi.org/10.1109/ACCESS.2020.3033789}.}}

\markboth
{Author \headeretal: Preparation of Papers for IEEE TRANSACTIONS and JOURNALS}
{Author \headeretal: Preparation of Papers for IEEE TRANSACTIONS and JOURNALS}

\corresp{Corresponding author: H\'{e}ctor M. Becerra (e-mail: hector.becerra@cimat.mx).}

\begin{abstract}
This paper addresses the distributed prescribed-time leader-following consensus problem for a class of high-order multi-agent systems (MASs) with perturbed nonlinear agents dynamics and where the topology of the network contains a directed spanning tree, with the leader as the root. Prescribed-time consensus means that an agreement state of the MAS is achieved in a preset time, introduced as a parameter of the control law, and this constant settling time is achieved independently of the agents' initial state. The proposed control method exhibits three main advantages: first, to our best knowledge, it is the first time that prescribed-time convergence in a consensus problem is achieved for agents with high-order nonlinear dynamics, using a robust leader-following protocol, which allows an effective rejection of matched disturbances in the agents' model. Second, the proposed controller provides control signals of lower magnitude than existing approaches. Third, the proposed consensus protocol does not have parameters to be adjusted depending on the connectivity of the considered communication graph.
\end{abstract}


\titlepgskip=-15pt

\maketitle

\section{Introduction}
Cooperative control of Multi-Agent Systems (MASs) is a broad topic involving many different related research problems, such as consensus, formation control, flocking, coverage control, among others; attracting considerable attention over the last decades due to their broad applications in different research areas (\cite{Y.Chen_J.Lu_X.Yu_D.J.Hill,W.Ren_Y.Cao}). 
A fundamental problem in cooperative control of MASs is to design distributed consensus protocols to make the autonomous agents to agree on some variable of interest. In this problem, each agent applies a controller that only uses information obtained from local interactions between neighboring agents.

A great deal of work has been recently published addressing the distributed consensus problem with real-time constraints, proposing distributed consensus algorithms with \textit{finite-time and fixed-time convergence}, see e.g.~\cite{AldanaConsensus2019,Ning2018}. As interesting examples of finite-time control of uncertain nonlinear systems, we can mention \cite{Mobayen-2017} for stabilization using nonsingular fast terminal sliding mode control, and \cite{Mobayen-2018} for synchronization of chaotic systems in the presence of external disturbances and time delays. In both finite and fixed-time convergence, the settling time is a finite value, but in the fixed-time convergence the settling time is uniformly bounded, meaning that the system converges to its equilibrium before an estimated bound that is independent of the initial state. Nevertheless, in the existing approaches of finite and fixed-time consensus, the convergence bound estimates may be too conservative, leading to over-engineering the system to satisfy real-time constraints~(\cite{Z.Zuo_Q.L.Han_B.Ning_X.Ge_X.M.Zhang}), resulting in large control efforts as a drawback.

In this work, we address the problem of designing a distributed control protocol to achieve consensus tracking of a leader in prescribed time for high-order MASs with nonlinear dynamics affected by disturbances. Unlike finite-time and fixed-time consensus, \textit{prescribed-time consensus} means that an agreement state of the MAS is accurately achieved at a pre-specified time, introduced as a parameter of the protocol, and this constant settling time is achieved independently of the agents' initial state, with no slack between the real and the desired settling time. Prescribed-time allows to efficiently schedule and execute tasks in a mission to be performed by a MAS, i.e., the agents in the MAS can reach the same state at exactly a preset-time provided by the schedule, then they can start the following task at a preset time, e.g. the efficient execution of a schedule for a team of robots performing tasks with temporospatial constraints (\cite{Gombolay-2018}), where efficient means that there is no time-outs of the robots between tasks, avoiding thus unnecessary large control efforts. These aspects are the main motivation to address the considered problem.

We consider a leader-following configuration where the topology among the followers and the leader contains a directed spanning tree, with the leader as the root. On the one hand, the importance of this consensus tracking problem is that the hard time constraint guaranteed by the prescribed-time convergence makes straightforward for a user to set the MAS convergence time and increases the potential of the engineering applications of consensus. On the other hand, the applicability of the leader-following scheme is larger than the leaderless case, since the goal is clearly defined by the leader, in contrast to the leaderless case where the consensus state results from the local interactions. The  problem  that  we address is very novel and challenging, with some efforts to solve it in the last years. Proof of it is that most of the existing results achieve just fixed-time convergence (~\cite{Z.Zuo_B.Tian_M.Defoort_Z.Ding, B.Tian_Z.Zuo_H.Wang,Zhang-2019, You-2020,Shi-2020, Zuo-2019, YangH-2020}) and only few works address prescribed-time convergence (~\cite{Y.Zhao_Y.Liu_G.Wein_W.Ren_G.Chen,Y.Zhao_Y.Liu_a,Y.Wang_Y.Song, Lu-2020}), since the high-order dynamics makes the problem more difficult. An important advantage of our approach, with respect to the referred fixed-time consensus protocols, is that we can preset the convergence time as a parameter of the control law. Regarding the few existing results on prescribed-time consensus of high-order systems, our approach has the main advantages of being robust against disturbances and generating lower control efforts.

We introduce a novel methodology to solve the prescribed-time consensus problem for a MAS of high-order dynamics. Our methodology consists in the tracking of suitable reference signals by using feedback controllers. The references are characterized by time base generators (TBGs), which are continuous time-dependent polynomial functions that converge to zero in a specified time~(\cite{H.M.Becerra_C.R.Vazquez_G.Arechavaleta_J.Delfin,Morasso1997}). Once these time-varying functions are designed, one of the main difficulties to solve the defined problem is to guarantee robust tracking of the reference signals defined for high-order systems to achieve accurate convergence of the consensus error in a preset time regardless of the interaction between agents and disturbances. We propose two consensus protocols in which only a leader agent gives the reference to the high-order MAS assuming that the topology of the network contains a directed spanning tree, with the leader as the root. The first one is a linear control protocol where feedback of the tracking error of the high-order TBG trajectories between neighbor agents is used. The second one is a robust consensus protocol, based on a sliding mode controller to provide closed-loop stability and robustness against disturbances. The proposed protocols can be applied to nonlinear high-order MAS that can be transformed to the normal form (e.g. the Brunovsky's canonical form) by state feedback linearization~(\cite{H.K.Khalil_J.W.Grizzle}), in which agents dynamics are represented as a chain of integrators. Convergence at the prescribed-time and global closed-loop stability are demonstrated theoretically and illustrated through simulations. 

In summary, the main contributions of the paper are: 
\begin{enumerate}
    \item A consensus approach for MASs with high-order agents is proposed, which ensures accurate convergence in a preset exact time independently of the initial conditions. 
    \item A robust consensus protocol is proposed based on a robust tracking controller, which guarantees convergence in a preset time regardless of the presence of unknown matched disturbances.
    \item Comparisons between our proposal and existing prescribed-time protocols are provided through simulations, showing that the proposed approach generates lower control efforts.
\end{enumerate}

An additional advantage of our approach is that the proposed control protocols do not use information about the network's connectivity, as used by other consensus approaches, which represents a robustness property that allows our protocols to work properly for either small or large number of agents without readjusting controller parameters.

\subsection{Related Work}

To date, few contributions addressing the prescribed-time consensus problem have been reported in the literature, mainly focusing on MASs with first-order and second-order agents, as all the references described in this paragraph. A class of distributed linear protocols were developed for linear MASs over both undirected and directed communication networks~(\cite{C.Yong_X.Guangming_L.Huiyang,C.Yong_X.Guangming_L.Huiyang_b,Y.Wang_Y.Song_D.J.Hill_M.Krstic,C.Liu_Q.Zhou_Y.Liu}). By using time-varying control gains, the agents in the network are forced to reach consensus at any preset time from any initial condition. Following another approach, the prescribed-time consensus problem has been transformed into a motion planning problem in which the developed consensus protocols are based on a time-varying sampling sequence convergent to an off-line desired settling time~(\cite{Y.Liu_Y.Zhao_W.Ren_G.Chen,Y.Zhao_Y.Liu,Y.Zhao_Y.Liu_b,Y.Liu_Y.Zhao_Z.Shi_D.Wei}). With a pre-specified settling
time, these protocols solve the consensus problem of linear
MASs over undirected and directed topologies, and directed switching topologies. In \cite{Ni-2020}, consensus tracking is investigated for second order MASs; the settling time bound is related to tunable parameters but it is not constant for all initial conditions.

Time-varying gains derived from TBGs have been used to solve different consensus problems with prescribed convergence for agents with first order dynamics (\cite{Kan2017,Yucelen2018,Ning2019}). In particular, the rendezvous problem~(\cite{Kan2017}) and the leader-following consensus problem (\cite{Yucelen2018,Ning2019}) have been addressed. The consensus algorithms with time-varying gains present important drawbacks, for instance, the time-varying gain becomes singular as the prescribed time is reached (\cite{C.Yong_X.Guangming_L.Huiyang,Kan2017,Yucelen2018}) or the time-varying gain is piecewise constant~(\cite{Y.Liu_Y.Zhao_W.Ren_G.Chen}) with Zeno behavior~(\cite{Zhang2001}). Commonly, a parameter is added to the controller to avoid the singularity of the time-varying gain. Unfortunately, with such modification, consensus cannot be reached in a constant time, but all agents arrive at a value around the consensus state in fixed-time. Following with MASs of first-order dynamics, TBGs have also been used in \cite{J.A.Colunga_C.R.Vazquez_H.M.Becerra_D.Gomez-Gutierrez,J.A.Colunga_C.R.Vazquez_H.M.Becerra_D.Gomez-Gutierrez_b} to impose a reference profile of a consensus error and consequently achieve prescribed-time convergence. In those works, it was introduced a prescribed-time distributed consensus protocol that requires to give the consensus value to each agent as a parameter of the controller. Under the same idea, in \cite{Ning-2020}, TBGs have been used to solve the bipartite consensus tracking in a preset time for second order agents.

Concerning consensus of high-order MASs, asymptotic convergence has been studied, e.g. in \cite{shao2018leader}. There exist recent results for consensus of high-order MAS based on finite-time control (\cite{S.Mondal_R.Su_L.Xie, Sakthivel-2019}) and fixed-time control (\cite{Z.Zuo_B.Tian_M.Defoort_Z.Ding, B.Tian_Z.Zuo_H.Wang,Zhang-2019, You-2020,Shi-2020, Zuo-2019, YangH-2020}). All of them are autonomous protocols not able to achieve a constant convergence time and only an overestimation of it is possible. In a different kind of approach, a distributed adaptive learning control is proposed in \cite{Yang-2020} for the coordination of high-order systems, but the convergence time cannot be preset. To our best knowledge, only few works address consensus in prescribed-time for high-order MASs~(\cite{Y.Zhao_Y.Liu_G.Wein_W.Ren_G.Chen,Y.Zhao_Y.Liu_a,Y.Wang_Y.Song, Lu-2020}). A specified-time consensus protocol in a leaderless scheme has been developed in~\cite{Y.Zhao_Y.Liu_G.Wein_W.Ren_G.Chen} for MASs with general linear dynamics over directed graphs containing a directed spanning tree and based on a motion planning strategy. A multi-leader approach has been presented in~\cite{Y.Zhao_Y.Liu_a}, where the followers are driven to the convex hull spanned by the leaders at a specified settling time if the undirected fixed topology is connected. A distributed time-varying control approach in a leader-following consensus scheme based on a finite-time observer has been investigated in~\cite{Y.Wang_Y.Song}, demonstrating consensus in a pre-specified finite-time under fixed directed topologies having a spanning tree. That work was extended for a class of linear unperturbed MASs by using event-triggered control in \cite{Lu-2020}.

To the authors' opinion, only the works~\cite{Y.Zhao_Y.Liu_G.Wein_W.Ren_G.Chen} and~\cite{Y.Wang_Y.Song} represent the closest approaches to the problem addressed in this paper. Nevertheless, in contrast to them, our approach presents several improvements: accurate convergence of the consensus error to zero in a preset time is guaranteed for the case where agents are affected by disturbances; neither Zeno behavior is produced nor singularities occur in the control signal, avoiding thus the main drawbacks of consensus approaches based on time-varying gains. In addition, it is shown that our consensus protocol can produce continuous control signals of lower magnitude than those approaches when a continuous auxiliary control is used. This is more suitable for certain applications, e.g. in formation control of MASs, where the consensus signal provides velocity references to be tracked by the agents~(\cite{Oh2015}). Finally, those existing approaches require to readjust some controller parameters depending on the algebraic connectivity of the network~(\cite{AldanaConsensus2019}), which is not the case of our approach.

The rest of this paper is organized as follows. Section \ref{sec:concepts} recalls basic concepts. Section \ref{sec:definitions} introduces the model of the MAS and defines the prescribed-time consensus problem. In Section \ref{sec:protocols}, a linear-feedback controller and a robust controller are proposed for the prescribed-time consensus problem in a leader-following scheme. Section \ref{sec:simulations} reports simulations of the proposed approach and comparisons with other protocols in the literature. Finally, Section \ref{sec:Conclusiones} presents some conclusions.

\section{Theoretical Preliminaries}\label{sec:concepts}

Let us first introduce some notation. $\bld{I}_n$ represents the identity matrix of dimension $n \times n$, $\bld{0}_{n \times n}$ denotes the zero matrix of dimension $n\times n$ . Let $\bld{1}_n$ and $\bld{0}_n$ be the $n$-dimensional column vectors with all entries equal to one and zero, respectively. $\bld{A} \otimes \bld{B}$ denotes the Kronecker product of matrices $\bld{A}$ and $\bld{B}$. 

\subsection{Time Base Generators}

\textit{Time base generators}~(TBGs) are parametric functions of time, particularly designed to drive a system in such a way that its state describes a convenient transient profile. TBGs have been previously used in \cite{H.M.Becerra_C.R.Vazquez_G.Arechavaleta_J.Delfin} to achieve prescribed-time convergence of a single first-order or high-order system. 

\begin{definition}(\cite{H.M.Becerra_C.R.Vazquez_G.Arechavaleta_J.Delfin})
A TBG of order $r$ and settling time $t_f$ is defined as a continuous and differentiable time-dependent function $h(t)$, described as~
\begin{equation}
h(t)=\left\{\begin{array}{ll}
\bld{\tau}(t)\cdot \bld{c} \quad &\text{if }t\in[0,t_f]
\\
0&\text{otherwise},
\end{array}\right.
\label{eq:TBG_hs}
\end{equation}
where $\bld{\tau}(t)=[t^{r}, \ t^{r-1}, \ \hdots, \ t, \ 1]$ is the time basis vector and $\bld{c}$ is a vector of coefficients of proper dimensions. 
\end{definition}

\begin{definition}
\label{def:tbgs} (\cite{H.M.Becerra_C.R.Vazquez_G.Arechavaleta_J.Delfin})
Consider a control system of order $n$. For the design of a prescribed-time controller, a collection of $n$ TBGs of order $r \geq 2n + 1$ is designed, fulfilling the following conditions at initial time and settling time $t_f$
\begin{align}\label{hfunction}
\forall k &\in  \{1,\dots,n\},\quad \forall j  \in \{0,\dots,n\}, \nonumber \\
h_k^{(j)}(t)|_{t=0} &= \left\{
\begin{array}{ll} 
1 & \textrm{ if } j = k-1 \\
0 & \textrm{ otherwise }
\end{array} \right. 
\\
h_k^{(j)}(t)|_{t\geq t_f} &= 
0, \nonumber 
\end{align}
where $h_k(t)=\bld{\tau}(t)\cdot \bld{c}_k$ denotes the $k$-th TBG in the collection for $t\in[0,t_f]$, and $h_k^{(j)}(t)$ denotes its $j$-th time derivative. \\
\end{definition}

The TBGs and their derivatives are grouped in a time-varying matrix as
\begin{equation}
\bld{H}(t) 
= 
\left[
\begin{array}{cccc}
h_1(t) & h_2(t) & \dots &h_n(t) \\
\dot{h}_1(t) & \dot{h}_2(t)& \dots & \dot{h}_n(t)\\
\vdots& \vdots& &\vdots\\
h^{(n-1)}_1(t)&h^{(n-1)}_2(t)& \dots&h^{(n-1)}_n(t)
\end{array}
\right],
\label{defH}
\end{equation}
then $\bld{H}(0) = \bld{I}_n$, and $\bld{H}(t\geq t_f) = \bld{0}_{n\times n}$ according to the constraints \eqref{hfunction}. 

Prescribed-time controllers for a single high-order system have been introduced in \cite{H.M.Becerra_C.R.Vazquez_G.Arechavaleta_J.Delfin}, using time-varying controllers based on the TBGs. Therein, the TBGs were used as time-varying gains as well as reference trajectories to be tracked. See \cite{H.M.Becerra_C.R.Vazquez_G.Arechavaleta_J.Delfin} for the details to calculate the coefficients $\bld{c}_k$ fulfilling the required constraints. For the work herein, we were inspired by the results of \cite{H.M.Becerra_C.R.Vazquez_G.Arechavaleta_J.Delfin}; the extension to a consensus protocol for a MAS is not trivial due to the interaction of the agents states in the communication topology. 

\subsection{Algebraic Graph Theory}

In a MAS, the communication between agents is represented by a graph. Let us recall some basic definitions on graph theory that can be found in \cite{W.Yu_G.Wen_G.Chen_J.Cao} and \cite{Z.Li_Z.Duan}.

\begin{definition} (\cite{W.Yu_G.Wen_G.Chen_J.Cao}) 
A communication graph is a tuple $\mathcal{G} = (\mathcal{V},\mathcal{E},\mathcal{A})$ that consists of a set of vertices representing agents $\mathcal{V} = \{\mathsf{v}_1,\dots,\mathsf{v}_N\}$, a set of edges representing communication channels $\mathcal{E} \subseteq \mathcal{V} \times \mathcal{V}$, and a weighted adjacency matrix $\mathcal{A} = [a_{ij}] \in \mathbb{R}^{N \times N}$ with non-negative entries $a_{ij}$, in particular, $a_{ij}>0$ if $(\mathsf{v}_i,\mathsf{v}_j)\in \mathcal{E}$ and $a_{ij} = 0$ if $(\mathsf{v}_i,\mathsf{v}_j) \notin \mathcal{E}$. The set of neighbors of agent $i$ is denoted by $N_i = \{j \in \mathcal{V} : (\mathsf{v}_j, \mathsf{v}_i) \in \mathcal{E} \}$.
\end{definition}

\begin{definition} (\cite{W.Yu_G.Wen_G.Chen_J.Cao})
A graph $\mathcal{G}$ is called \emph{directed} if the edges are ordered pairs, i.e., $(\mathsf{v}_i,\mathsf{v}_j)$ and $(\mathsf{v}_j,\mathsf{v}_i)$ denote different edges. A graph $\mathcal{G}$ is called \emph{undirected} if the edges are unordered pairs, i.e., $(\mathsf{v}_i,\mathsf{v}_j)$ and $(\mathsf{v}_j,\mathsf{v}_i)$ denote the same edge.
\end{definition} 

In the MAS framework, $(\mathsf{v}_i,\mathsf{v}_j)$ denotes that agent $\mathsf{v}_j$ can obtain information from agent $\mathsf{v}_i$.

\begin{definition} (\cite{W.Yu_G.Wen_G.Chen_J.Cao})
A \emph{directed path} from vertex $\mathsf{v}_i$ to $\mathsf{v}_j$ is a sequence of distinct vertices $\mathsf{v}_{i}$, $\mathsf{v}_{i_1}$,..., $\mathsf{v}_{i_r}$, $\mathsf{v}_{j}\in\mathcal{V}$ and edges $(\mathsf{v}_i,\mathsf{v}_{i_1}), (\mathsf{v}_{i_1},\mathsf{v}_{i_2}), \dots, (\mathsf{v}_{i_r},\mathsf{v}_j)\in \mathcal{E}$. 

A graph $\mathcal{G}$ is said to be \emph{connected} if there exists an undirected path between any two distinct vertices $\mathsf{v}_i$ and $\mathsf{v}_j$ in $\mathcal{V}$.

A directed graph $\mathcal{G}$ is said to have a \emph{directed spanning tree} if $\mathcal{G}$ has at least one vertex $\mathsf{v}_i$, named \emph{root}, such that for any other vertex $\mathsf{v}_j\in\mathcal{V}\setminus\{\mathsf{v}_i\}$ there is a directed path from $\mathsf{v}_{i}$ to $\mathsf{v}_{j}$. It is assumed that no self-loops exist in the considered graphs.
\end{definition}

\begin{definition} (\cite{W.Yu_G.Wen_G.Chen_J.Cao}) Let $\mathcal{G}$ be a graph with $N$ vertices. The Laplacian matrix of $\mathcal{G}$ is defined as the $N \times N$ matrix $\bld{L} = [l_{ij}]$, where
\begin{align}
l_{ij}= 
\left\{
\begin{array}{ll}
-a_{ij}, & \mathrm{ if }\quad i \neq j, \\
\sum\limits_{k=1, k \neq i}^{N} a_{ik}, & \mathrm{ if }\quad i = j.
\end{array}\right.
\label{laplacian_matrix}
\end{align}
\end{definition}

\section{MAS Definition and Problem Statement}\label{sec:definitions}

Our setup consists of a multi-agent system (MAS) formed by a collection of $N$ agents named \emph{followers} whose dynamics are described by nonlinear systems with relative degree $n$, an agent named \emph{leader}, and a communication graph $\mathcal{G}$ with $N+1$ vertices, each one associated to a different agent. It is assumed that the model of each follower agent is in the \textit{normal form}~(\cite{H.K.Khalil_J.W.Grizzle}), e.g. for the $i$-th agent
\begin{equation}
\begin{array}{l}
\dot{x}_{i1}=x_{i2},\\
\qquad\vdots\\
\dot{x}_{i(n-1)}=x_{in},\\
\dot{x}_{in}=f_i(\bld{x}_i,\bld{\varphi}_i) + g_i(\bld{x}_i,\bld{\varphi}_i) u_i(t) +\rho_i(t),\\
\dot{\varphi}_{i1}=q_{i1}(\bld{x}_i,\bld{\varphi}_i),\\
\qquad\vdots\\
\dot{\varphi}_{ir}=q_{ir}(\bld{x}_i,\bld{\varphi}_i),\\
y_i=x_{i1},
\end{array}\label{chainInt_nonlinear}
\end{equation}
where $\bld{x}_i = [x_{i1},\dots,x_{in}]^T \in \mathbb{R}^n$ is the agent's state, $u_i(t) \in \mathbb{R}$ is the agent's control input, $f_i(\bld{x}_i,\bld{\varphi}_i)$ and $g_i(\bld{x}_i,\bld{\varphi}_i)$ are smooth nonlinear functions, $\rho_i(t) \in \mathbb{R}$ represents bounded lumped uncertainties that include disturbances and nonlinear uncertainties (i.e. $\rho_i(t)=\rho_i'(t)+\Delta f_i(\bld{x}_i,\varphi_i)$, where $\rho_i'(t)$ are time-varying bounded disturbances and $\Delta f_i(\bld{x}_i,\varphi_i)$ are bounded unknown nonlinearities), $\bld{\varphi}_i = [\varphi_{i1},\dots,\varphi_{ir}]^T \in \mathbb{R}^r$ is the agent's zero dynamics, and $y_i \in \mathbb{R}$ is the agent's output. 
It is assumed that the relative degree is well defined ($g_i(\bld{x_i},\bld{\varphi}_i) \neq 0$), and the zero dynamics is stable.

Furthermore, the leader is an agent whose dynamics are given as an integrator chain, i.e.
\begin{align}
\dot{x}_{lk} &= x_{l(k+1)}, \quad k = 1, \dots,n-1\nonumber \\
\dot{x}_{ln} &= u_{l}(t), 
\label{dyn_leader}
\end{align}
where $\bld{x}_l(t) \in \mathbb{R}^n$ is the leader's state and $u_{l}(t) \in \mathbb{R}$ is the leader's control input.

The input-output dynamics \eqref{chainInt_nonlinear} of each $i$-th follower agent can be conveniently represented as an $n$-integrators chain as
\begin{align}
\dot{x}_{ik} &= x_{i(k+1)}, \quad k = 1, \dots,n-1 \nonumber \\
\dot{x}_{in} &= v_i(t) +\rho_i(t),\label{chain_integrators}
\end{align}
by applying the control input
\begin{equation}\label{eq:CLtransformation}
    u_i = (-f_i(\bld{x}_i,\bld{\varphi}_i) + v_i(t))/g_i(\bld{x}_i,\bld{\varphi}_i),
\end{equation} 
where  $v_i(t) \in \mathbb{R}$ is an auxiliary control input. Then the $i$-th agent's dynamic \eqref{chain_integrators} can be expressed in a vectorial form as
\begin{align} \label{xiDynamic}
\dot{\bld{x}}_i &=\bld{Ax}_i + \bld{B}(v_i(t) + \rho_i(t)), \quad i \in \{1,\dots,N\}
\end{align}
with adequate constant matrices $\bld{A} \in \mathbb{R}^{n\times n}$ and $\bld{B} \in \mathbb{R}^{n}$, where $(\bld{A},\bld{B})$ is controllable. Similarly, the leader's dynamics \eqref{dyn_leader} can be written as
\begin{align}
\dot{\bld{x}}_l = \bld{A}\bld{x}_l + \bld{B}u_l(t),
\label{leader_System}
\end{align}

The class of agents that can be represented in the form of \eqref{chain_integrators} is broad. In particular, any SISO linear time-invariant system $\dot{\bld{x}}_i = \bar{\bld{A}}\bld{x}_i + \bar{\bld{B}}(u_i(t) + \rho_i(t))$ can also be transformed into an $n$-order integrator system \eqref{chain_integrators} provided it is controllable, observable and has no transmission zeros, by transforming the system into the so-called \textit{observability canonical form}~(\cite{T.Kailath}) and applying an input that cancels the open-loop dynamics of the $n$-th state equation.

Now, let us introduce the concept of consensus error in the leader-following scheme.

Consider a MAS, where each follower is already in its integrator chain form \eqref{chain_integrators}, i.e., the control law \eqref{eq:CLtransformation} is applied to each follower agent. The \emph{consensus error} for each $i$-th follower is given by
\begin{align}
\bld{e}^{f}_{i}(t)&= \left[ e_{i1},\dots, e_{in} \right]^T \nonumber \\
&= \sum_{j \in N_i\setminus \{l\}} a_{ij} \left(\bld{x}_{j}(t)-\bld{x}_{i}(t)\right)-b_i\left(\bld{x}_{i}(t)-\bld{x}_{l}(t)\right),
\label{follower_error}
\end{align}
where $a_{ij}$ are the entries of the graph adjacency matrix, $N_i\setminus \{l\}$ denotes the set of neighbors of agent $i$-th excepting the leader, and $b_i = a_{il}$ represents the adjacency to the leader ($b_i = a_{il} > 0$ if agent $i$ is a neighbor of the leader, $b_i = a_{il} = 0$ otherwise).\\

\label{def:problem}
\noindent \textbf{Problem statement.}
Consider a MAS and assume that the control law \eqref{eq:CLtransformation} is applied to each agent. The prescribed-time consensus problem consists in designing a protocol in the form $v_i=\eta_i(\bld{e}^f_i,t)$ for each follower agent, such that the state of all the agents reach a consensus state, given by the leader's state $\bld{x}_l(t)$, in a prescribed time $t_f$ from any initial state $\bld{x}_i(0)$, i.e., $\forall i \in \{1,\dots,N\}$, $\bld{x}_i(t) \to \bld{x}_l(t)$ as $t \to t_f$. 

\remark{Notice that we consider that the leader’s state $\bld{x}_l(t)$, defining the consensus state, can be constant or time-varying. In the last case, $\bld{x}_l(t)$ can be generated through a no null initial condition of the leader's state $\bld{x}_l(0)$ in the high-order variables or by introducing a control input to the leader $u_l(t)$.}

\begin{assumption}
The topology of the communication graph $\mathcal{G}$ is a directed graph that has a spanning tree in which the leader acts as the root. The leader vertex will be denoted as $l$.
\label{assumption_1}
\end{assumption}


\section{Prescribed-Time Consensus Protocols}\label{sec:protocols}

In this section, two protocols for the prescribed-time consensus problem are introduced. First, a linear feedback-based consensus protocol is presented. Later, a consensus protocol based on a sliding mode controller is proposed, providing robustness against disturbances while maintaining the prescribed-time convergence property.

Before introducing the protocols, let us first demonstrate that  Assumption \ref{assumption_1} implies that the consensus is reached when the consensus errors are null for all the agents. 

\begin{lemma}\label{lemma}
Consider a high-order MAS modeled as in Section \ref{sec:definitions}, fulfilling Assumption \ref{assumption_1}, and the control law \eqref{eq:CLtransformation} for each follower agent. If for each $i$-th follower $\bld{e}_i^f(t)=\mathbf{0}$, then $\bld{x}_i(t)=\bld{x}_l(t)$, i.e., consensus is reached.
\end{lemma}

\begin{IEEEproof}
Let us define $\bld{M} = diag (b_1,\dots, b_N)$ (a diagonal matrix with entries $b_1,\dots, b_N$) and $\bld{m} = \bld{M}\cdot\bld{1}_N = [b_1,\dots,b_N]^T$. Moreover, let us denote $\bld{x}(t)=[\bld{x}^T_1(t),...,\bld{x}^T_N(t)]^T$ and $\bld{e}^f(t)=[{\bld{e}^{fT}_1}(t),...,{\bld{e}^{fT}_N}(t)]^T$. By using the Laplacian matrix $\bld{L}$ and \eqref{follower_error}, $\bld{e}^f(t)$ can be expressed as
\begin{align}\label{eq-aux}
  \bld{e}^f(t)&=-\left(\bld{L}\otimes\bld{I}_n\right)\bld{x}(t)-\left(\bld{M}\otimes\bld{I}_n\right)\left(\bld{x}(t)-\left(\bld{1}_N\otimes\bld{x}_l(t)\right)\right)\nonumber\\
&=-\left( \bld{L}\otimes\bld{I}_n+\bld{M}\otimes\bld{I}_n\right)\bld{x}(t)+\left(\bld{M}\cdot\bld{1}_N\otimes\bld{x}_l(t)\right)\nonumber\\
&=-\left(\bld{L}\otimes\bld{I}_n+\bld{M}\otimes\bld{I}_n\right)\bld{x}(t)+\left(\bld{m}\otimes\bld{x}_l(t)\right)\nonumber\\
&=-\left(\bld{L}\otimes\bld{I}_n+\bld{M}\otimes\bld{I}_n\right)\bld{x}(t)+\left(\bld{m}\otimes\bld{I}_n\right)
\left(1\otimes\bld{x}_l(t)\right).
\end{align}

Lemma 1 from~\cite{shao2018leader} for high-order systems ensures $\left( \bld{L} \otimes \bld{I}_n + \bld{M} \otimes \bld{I}_n \right)^{-1}\left(\bld{m} \otimes \bld{I}_n \right) = \bld{1}_N \otimes \bld{I}_n$, which holds by Assumption \ref{assumption_1}. By using this result and the hypothesis $\bld{e}^f_i(t)=\bld{0}_n$, the equation \eqref{eq-aux} can be solved for $\bld{x}(t)$ as
\begin{align}
\bld{x}(t)&=\left(\bld{L}\otimes\bld{I}_n+\bld{M}\otimes\bld{I}_n\right)^{-1}\left(\bld{m}\otimes\bld{I}_n\right)\left(1\otimes\bld{x}_l(t)\right)\nonumber\\
&=\left(\bld{1}_N\otimes\bld{I}_n\right)\left(1\otimes\bld{x}_l(t)\right)\nonumber\\
&=\bld{1}_N\otimes\bld{x}_l(t).
\end{align}

Therefore, consensus of the high-order MAS in the leader-following scheme is achieved when, for each $i$-th follower, $\bld{e}_i^f(t)=\mathbf{0}_n$.
\end{IEEEproof}

Now, in order to solve the prescribed-time consensus problem, our approach is to design a protocol $v_i$, for each $i$-th agent, that enforces the consensus error's transient behavior $\bld{e}^{f*}_{i}(t)=\bld{H}(t)\bld{e}^f_i (0)$, named the \emph{TBG reference trajectory} for the $i$-th follower agent, where $\bld{H}(t)$ is defined as in \eqref{defH} and $\bld{e}^f_i(0)$ denotes the initial consensus error. In this context, the \emph{tracking error} for the $i$-th agent is defined as
\begin{align}
\bld{\xi}_i(t) = 
\left[ \xi_{i1},\dots, \xi_{in} \right]^T = \bld{e}^{f}_{i}(t) - \bld{H}(t)\bld{e}^f_i (0).
\label{tracking_error_TBG_STC}
\end{align}

Thus, the consensus protocols to be defined for each agent, $v_i$, will enforce $\bld{\xi}_i(t)=\bld{0}_{n}$ $\forall t\geq0$. In fact, since $\bld{H}(t)=\bld{0}_{n\times n}$ $\forall t\geq t_f$, then $\bld{\xi}_i(t)=\bld{0}_{n}$ $\forall t\geq0$ implies $\bld{e}^{f}_{i}(t)=\bld{H}(t)\bld{e}^f_i (0)=\bld{0}_{n}$ $\forall t\geq t_f$, i.e., if the tracking error is null then the consensus error is null at $t_f$, consequently, if this occurs for all the agents then the consensus is reached at $t_f$.

\remark{It is possible to perfectly track the TBG reference trajectory from the initial time (i.e., $\bld{\xi}_i(t)=\bld{0}_{n}$ $\forall t\geq0$) due to its coherence with the initial consensus error and the error dynamics. In detail, notice that $\bld{H}(t)\bld{e}^{f}_{i}(0)=\bld{e}^f_i(0)$ at $t=0$, since $\bld{H}(0)=\bld{I}_{n}$ by Definition \ref{def:tbgs} and \eqref{defH}. Moreover, the definition of $\bld{H}(t)$ implies that $\dot{e}^{f}_{ij}(t)=e^{f}_{i(j+1)}(t)$ $\forall t>0$, $\forall j\in\{1,...,(n-1)\}$, i.e., the reference trajectory represents an $n$-integrator chain, similar to the dynamics of the followers and the leader.}

\subsection{Prescribed-Time Consensus with a Linear Protocol}\label{sec:protocols_lin}

The following theorem proposes a feedback-based protocol able to drive a high-order MAS to consensus in prescribed time, providing closed-loop stability of the tracking error.

\begin{theorem}
\label{theo_high_order_lin}
Consider a high-order MAS modeled as in Section \ref{sec:definitions} with $\rho_i=0$, fulfilling Assumption \ref{assumption_1}, and the control law \eqref{eq:CLtransformation} for each follower agent. Consider TBG functions for a system of order $n$ as in \eqref{hfunction}, gathered in the matrix $\bld{H}(t)$ as in \eqref{defH}, and define the time-varying gain vector $\bld{K}_t(t)=[h_1^{(n)}(t),\dots, h_n^{(n)}(t)]$. For each agent $i$, define $\beta_i = \left(b_i + \sum_{j \in N_i\setminus\{l\}}a_{ij}\right)$ and consider its tracking error \eqref{tracking_error_TBG_STC} and the vector $\bld{\xi}_{i2:in} 
= \left[\xi_{i2}, \dots, \xi_{in} \right]^T$.\\
Considering the linear controller defined for each follower agent $i$ as
\begin{equation}
v_i = \beta_i^{-1} \left( b_i u_l+ \sum_{j \in N_i\setminus\{l\}}a_{ij}v_j-\bld{K}_t(t) \bld{e}^f_i(0)+\bld{K}_{fr}\bld{\xi}_{i2:in} \right) \label{eq:controller_lin}
\end{equation}
there exist gains $\bld{K}_{fr}\in \mathbb{R}^{n-1}$ such that the agents' tracking errors $\bld{\xi}_i(t)$ are globally asymptotically stable. Furthermore, prescribed-time convergence of the followers' state $\bld{x}_{i}(t)$ to the leader's state $\bld{x}_{l}(t)$ is achieved at time $t_f$, independently of their initial states.

\end{theorem}

\begin{IEEEproof}
\emph{Part I.} First, let us prove closed-loop stability of the tracking error of each agent $\bld{\xi}_{i}$.

By taking the time derivative of the consensus error of the $i$-th follower \eqref{follower_error}, using the dynamics of the followers \eqref{xiDynamic} and the leader \eqref{leader_System}, and assuming $\rho_i=0$, the dynamics of the consensus error of the $i$-th follower is expressed as
\begin{align}
\dot{\bld{e}}^{f}_{i}(t) &=\bld{A}\bld{e}^f_i(t) +\bld{B}\Big(-\beta_i v_i+\sum_{j \in N_i}a_{ij}v_j+b_i u_l\Big).
\label{dot_follower_error_lin}
\end{align}

The time derivative of the tracking error \eqref{tracking_error_TBG_STC} requires to compute $\dot{\bld{H}}(t)$. By employing the matrices $\bld{A}$ and $\bld{B}$ in \eqref{xiDynamic}, we obtain
\begin{align*}
\bld{BK}_t(t) &=  \left[\begin{array}{c}0\\0\\ \vdots \\ 0 \\ 1\end{array}\right]  [h_1^{(n)}(t),h_2^{(n)}(t),\dots,h_n^{(n)}(t)] \\
&=\left[\begin{array}{cccc}0 & 0 &\dots &0\\
0 & 0& \dots & 0\\
\vdots& \vdots& \ddots &\vdots\\
0 & 0& \dots & 0\\
h^{(n)}_1(t)&h^{(n)}_2(t)& \dots&h^{(n)}_n(t)
\end{array}
\right],\nonumber
\end{align*}
and
\begin{align*}
&\bld{AH}(t)= \\ 
&=\tiny{
\left[\begin{array}{ccccc}0 & 1 & 0 & \dots & 0\\
0 & 0 & 1 & \dots & 0\\
\vdots& \vdots& \vdots &\ddots &\vdots\\
0 & 0 & 0 & \dots & 1\\
0 & 0 & 0 & \dots &0 \end{array}\right] 
\left[\begin{array}{cccc}h_1(t) & h_2(t) & \dots &h_n(t)\\
\dot{h}_1(t) & \dot{h}_2(t)& \dots & \dot{h}_n(t)\\
\vdots& \vdots& &\vdots\\
h^{(n-1)}_1(t)&h^{(n-1)}_2(t)& \dots&h^{(n-1)}_n(t)\end{array}\right]} \\ & = 
\left[\begin{array}{cccc}\dot{h}_1(t) & \dot{h}_2(t)& \dots & \dot{h}_n(t)\\
\vdots& \vdots& \ddots &\vdots\\
h^{(n-1)}_1(t)&h^{(n-1)}_2(t)& \dots&h^{(n-1)}_n(t)\\
0 & 0 & \dots &0 \\
\end{array}\right].
\end{align*}

Taking the time derivative of \eqref{defH}, it results 
\begin{align*}
\dot{\bld{H}}(t)&= \left[\begin{array}{cccc}\dot{h}_1(t)&\dot{h}_2(t)&\dots&\dot{h}_n(t) \\
\ddot{h}_1(t) & \ddot{h}_2(t)& \dots & \ddot{h}_n(t)\\
\vdots& \vdots& &\vdots\\
h^{(n)}_1(t)&h^{(n)}_2(t)& \dots&h^{(n)}_n(t)
\end{array}
\right].\nonumber
\end{align*}

Then, we readily derive that
\begin{align}
\dot{\bld{H}}(t)  &= \bld{AH}(t) + \bld{BK}_t(t).
\label{formula_BKt}
\end{align}

Then, using \eqref{dot_follower_error_lin} and 
\eqref{formula_BKt}, the time derivative of the tracking error \eqref{tracking_error_TBG_STC} can be expressed as
\begin{align*}
\begin{array}{l}
\dot{\bld{\xi}}_i(t)=\dot{\bld{e}}^{f}_{i}(t)-\dot{\bld{H}}(t)\bld{e}^f_i(0)=\\
\,\bld{A}\bld{\xi}_i(t)+\bld{B}\Big(-\beta_i v_i+\sum\limits_{j \in N_i\setminus\{l\}}a_{ij}v_j+b_i u_l-\bld{K}_t(t) \bld{e}^f_i (0)\Big)
\end{array}
\end{align*}

Given the canonical form of $\bld{A}$ and $\bld{B}$, the tracking error dynamics are represented as
\begin{align}
\dot{\xi}_{ik}(t)&= \xi_{i(k+1)},\quad k = 1,\dots,n-1, \nonumber \\
\dot{\xi}_{in}(t)&=-\beta_i v_i+\sum_{j \in N_i\setminus\{l\}}a_{ij}v_j+b_i u_l-\bld{K}_t(t) \bld{e}^f_i(0).
\label{tracking_error_integrator_dynamics_lin}
\end{align}

Substituting the control protocol \eqref{eq:controller_lin} into the above expression yields
\begin{equation}
\begin{array}{l}
\dot{\xi}_{ik}= \xi_{i(k+1)},\quad \text{for }k = 1,\dots,n-1,\\
\dot{\xi}_{in}=-\bld{K}_{fr}\bld{\xi}_{i2:in},
\end{array}
\label{eq:red_errsyst_lin}
\end{equation}
which can be enforced to exhibit global asymptotic stability through an appropriate choice of the gain vector $\bld{K}_{fr}$, since this dynamics can be seen as a controllable chain of $n-1$ integrators with a feedback input $-\bld{K}_{fr}\bld{\xi}_{i2:in}$. Then, for each $i$-th agent, $\bld{\xi}_i(t)$ is globally asymptotically stable. \\

\noindent \emph{Part II.}  Now let us prove prescribed-time consensus.

Recall that at initial time $t=0$, $\bld{H}(0)=\bld{I}_n$ and thus, the vector of tracking error for each agent initiates null, i.e., $\bld{\xi}_i(0)= \bld{0}_n$ according to \eqref{tracking_error_TBG_STC}. This implies that $\bld{\xi}_i(t)=\bld{0}_n ~ \forall t\geq 0$ by the stability of \eqref{eq:red_errsyst_lin}, since the agents of the MAS are not perturbed and the tracking error system starts at its equilibrium point. Thus, it follows that $\bld{e}^f_i(t)=\bld{H}(t)\bld{e}^f_i(0) ~ \forall t\geq 0$ in accordance to the definition of the tracking error \eqref{tracking_error_TBG_STC}. Finally, given that $\bld{H}(t_f)=\bld{0}_{n\times n}$, then $\bld{e}^f_i(t_f)=\bld{0}_n$. Since this occurs for all the agents, Lemma \ref{lemma} implies that consensus is reached at $t_f$. 
\end{IEEEproof}

\remark{Although the result in Theorem \ref{theo_high_order_lin} achieves the same control objective than the existing approaches \cite{Y.Zhao_Y.Liu_G.Wein_W.Ren_G.Chen,Y.Zhao_Y.Liu_a,Y.Wang_Y.Song, Lu-2020}, which are valid for unperturbed systems, our result has the advantage that smooth auxiliary control signals are generated. Due to the properties of the reference trajectories given by the TBGs, the auxiliary control signals start in zero, evolve in a smooth way, and vanish at the prescribed time of convergence, avoiding the large usual initial value of other control laws.}


\subsection{Robust Prescribed-Time Consensus}\label{sec:protocols_st}

The previous result can be extended to effectively deal with disturbances while achieving prescribed-time convergence, which is not achieved by the existing results in the literature. As explained before, the prescribed-time consensus problem can be solved by designing a protocol $v_i$ for each follower agent such that $\bld{\xi}_i(t)=\bld{0}_{n}$ $\forall t>0$. Since the tracking error exhibits high-order dynamics, in order to apply the sliding mode control technique, a \emph{sliding surface} (an algebraic variety in the state space containing the origin) is firstly designed in such a way that $\bld{\xi}_i(t)$ is asymptotically stable when confined to the sliding surface, later, a sliding mode control term is designed in order to maintain the tracking error on the surface. The sliding surface designed for the $i$-th follower agent is characterized by a variable $s_i(\bld{\xi}_i(t))$, in such a way that $s_i(\bld{\xi}_i(t))=0$ when the tracking error is evolving on the surface. In particular, we define
\begin{align}
s_i(t) = \left[ \bld{K}_{fr}, 1 \right] \bld{\xi}_i(t)= \bld{K}_{fr}\bld{\xi}_{i1:i(n-1)} + \xi_{in},
\label{sliding_surface}
\end{align}
where $\bld{K}_{fr} \in \mathbb{R}^{n-1}$ is a gain vector and $\bld{\xi}_{i1:i(n-1)} = \left[ \xi_{i1},\dots, \xi_{i(n-1)} \right]^T$.
\begin{theorem}
\label{theo_stc_high_order}
Consider a high-order perturbed MAS modeled as in Section \ref{sec:definitions}, fulfilling Assumption 1, and the control law \eqref{eq:CLtransformation} for each follower agent. Consider TBG functions for a system of order $n$ as in \eqref{hfunction}, gathered in the matrix $\bld{H}(t)$ as in \eqref{defH}, and define the time-varying gain vector $\bld{K}_t(t)=[h_1^{(n)}(t),\dots, h_n^{(n)}(t)]$. For each agent $i$, define $\beta_i = \left(b_i + \sum_{j \in N_i\setminus\{l\}}a_{ij}\right)$ and consider its tracking error \eqref{tracking_error_TBG_STC}, the vector $\bld{\xi}_{i2:in} 
= \left[\xi_{i2}, \dots, \xi_{in} \right]^T$ and the sliding surface \eqref{sliding_surface}.\\
Considering the nonlinear controller defined for each follower agent $i$ as
\begin{align}
v_i &= \beta_i^{-1} \left( \mu_i + \nu_i\right), \label{stc_control} \\
\mu_i &= b_i u_l+ \sum_{j \in N_i\setminus\{l\}}a_{ij}v_j-\bld{K}_t(t) \bld{e}^f_i(0)+\bld{K}_{fr}\bld{\xi}_{i2:in}, \nonumber \\
\nu_i&=k_1\sign(s_i), \nonumber
\end{align}
there exist gains $\bld{K}_{fr}\in \mathbb{R}^{n-1}$ and $k_1 >0$ such that the agents' tracking errors $\bld{\xi}_i(t)$ are globally asymptotically stable. Furthermore, prescribed-time convergence of the followers' state $\bld{x}_{i}(t)$ to the leader's state $\bld{x}_{l}(t)$ is achieved at time $t_f$, independently of their initial states.

\end{theorem}
\begin{IEEEproof}
\emph{Part I.} First, let us prove closed-loop stability of the tracking error of each agent $\bld{\xi}_{i}$.

By taking the time derivative of the consensus error of the $i$-th follower \eqref{follower_error} and using the perturbed dynamics of the followers \eqref{xiDynamic} and the leader \eqref{leader_System}, the dynamics of the consensus error of the $i$-th follower is expressed as
\begin{align}
\dot{\bld{e}}^{f}_{i}(t) = \bld{A}\bld{e}^f_i(t) +& \bld{B}\Big(-\beta_i v_i+\sum_{j \in N_i}a_{ij}v_j-\beta_i \rho_i \nonumber \\ +& \sum_{j \in N_i\setminus\{l\}}a_{ij}\rho_j+b_i u_l\Big). 
\label{dot_follower_error}
\end{align}

By employing the expression $\dot{\bld{H}}(t)= \bld{AH}(t) + \bld{BK}_t(t)$, derived in the proof of Theorem \ref{theo_high_order_lin}, and introducing \eqref{dot_follower_error}, the time derivative of the tracking error \eqref{tracking_error_TBG_STC} can be expressed as
\begin{align*}
\dot{\bld{\xi}}_i(t)&=\dot{\bld{e}}^{f}_{i}(t)-\dot{\bld{H}}(t)\bld{e}^f_i(0),\\
&=\bld{A}\bld{\xi}_i(t)+\bld{B}\Big(-\beta_i v_i+\sum_{j \in N_i\setminus\{l\}}a_{ij}v_j-\beta_i \rho_i \\
&+\sum_{j \in N_i\setminus\{l\}}a_{ij}\rho_j+b_i u_l-\bld{K}_t(t) \bld{e}^f_i (0)\Big).
\end{align*}

Given the canonical form of $\bld{A}$ and $\bld{B}$, the tracking error dynamics are represented as
\begin{align}
\dot{\xi}_{ik}(t)&= \xi_{i(k+1)},\quad k = 1,\dots,n-1, \nonumber \\
\dot{\xi}_{in}(t)&=-\beta_i v_i+\sum_{j \in N_i\setminus\{l\}}a_{ij}v_j-\beta_i \rho_i \nonumber \\
&+\sum_{j \in N_i\setminus\{l\}}a_{ij}\rho_j+b_i u_l-\bld{K}_t(t) \bld{e}^f_i(0).
\label{tracking_error_integrator_dynamics}
\end{align}

By using this expression, the dynamics of the variable $s_i(t)$ \eqref{sliding_surface} is computed as 
\begin{align*}
\dot{s}_i(t) = \bld{K}_{fr}\bld{\xi}_{i2:in}&-\beta_i v_i+\sum_{j \in N_i\setminus\{l\}}a_{ij}v_j-\beta_i\rho_i \\
&+\sum_{j\in N_i\setminus\{l\}}a_{ij}\rho_j+b_i u_l-\bld{K}_t \bld{e}^f_i(0).
\end{align*}

Substituting the control protocol \eqref{stc_control} into the above expression yields
\begin{align*}
\dot{s}_i&=-k_1\sign(s_i)-\beta_i\rho_i+\sum_{j \in N_i\setminus\{l\}}a_{ij} \rho_j. 
\end{align*}

Let us propose the following Lyapunov candidate function:
\begin{align*}
V_i=\frac{1}{2} s_i^2. 
\end{align*}

The time derivative of $\dot{V}_i$ along the trajectory of the sliding surface is:
\begin{align*}
\dot{V}_i=-k_1|s_i|+\left(\sum_{j \in N_i\setminus\{l\}}a_{ij} \rho_j-\beta_i\rho_i\right)s_i. 
\end{align*}

Then, we have that
\begin{align}
\dot{V}_i\leq-\left(k_1+\beta_i|\rho_i|-\sum_{j \in N_i\setminus\{l\}}a_{ij} |\rho_j|\right)|s_i| \leq 0, 
\label{eq:derLyap}
\end{align}
is fulfilled whenever
\begin{align}
k_1 \geq \beta_i|\rho_i|-\sum_{j \in N_i\setminus\{l\}}a_{ij} |\rho_j|. 
\label{eq:gainCondition}
\end{align}

If the condition \eqref{eq:gainCondition} on $k_1$ holds, the dynamics of the tracking error for each agent is constrained to the sliding surface, i.e., $s_i= \dot{s}_i=0$, regardless of the presence of disturbances. In such case, from \eqref{sliding_surface} it follows that
\begin{align*}
\xi_{in}=-\bld{K}_{fr}\bld{\xi}_{i1:i(n-1)}.
\end{align*}

Then, considering this expression and the first $n-1$ equations of the tracking error dynamics \eqref{tracking_error_integrator_dynamics}, the behavior of the tracking error on the sliding surface results in
\begin{equation}
\begin{array}{l}
\dot{\xi}_{ik}= \xi_{i(k+1)},\quad \text{for }k = 1,\dots,n-2,\\
\dot{\xi}_{i(n-1)}=\xi_{in} = -\bld{K}_{fr}\bld{\xi}_{i1:i(n-1)},
\end{array}
\label{eq:red_errsyst}
\end{equation}
which can be enforced to exhibit global asymptotic stability through an appropriate choice of the gain vector $\bld{K}_{fr}$. Then, for each $i$-th agent, $\bld{\xi}_i(t)$ is globally asymptotically stable. \\

\noindent \emph{Part II.}  Based on the result of Part I, now let us prove prescribed-time consensus.

At initial time $t=0$, $\bld{H}(0)=\bld{I}_n$ and thus the tracking error vector for each agent is initially null, i.e., $\bld{\xi}_i(0)= \bld{0}_n$ according to \eqref{tracking_error_TBG_STC}. This implies that $s_i(0)=0$ according to \eqref{sliding_surface}. Due to the stability of the sliding surface dynamics \eqref{eq:derLyap}, the system's trajectory never leaves the surface once it reaches $s_i=0$ (\cite{Utkin2013}), which implies that $s_i(t)=0$ $\forall t> 0$, regardless of the disturbances $\rho_i(t)\neq0$. Then, the tracking error vector keeps null all the time since it is constrained to the manifold $s_i=0$, i.e., $\bld{\xi}_i(t)=\bld{0}_n ~ \forall t\geq 0$ by the stability of \eqref{eq:red_errsyst}. Thus, it follows that $\bld{e}^f_i(t)=\bld{H}(t)\bld{e}^f_i(0) ~ \forall t\geq 0$ in accordance to \eqref{tracking_error_TBG_STC}. Finally, given that $\bld{H}(t_f)=\bld{0}_{n\times n}$, then $\bld{e}^f_i(t_f)=\bld{0}_n$. Since this occurs for all the agents, Lemma \ref{lemma} implies that consensus is reached at $t_f$.
\end{IEEEproof}

\subsection{Further Extension for Continuous Robust Control}

The proposed robust controller of Theorem~\ref{theo_stc_high_order} provides the theoretical properties to guarantee prescribed-time convergence of the MAS using discontinuous control signals. For some applications, the use of discontinuous control signals could be unsuitable. Nevertheless, a continuous control law may be formulated to achieve the same goal by using high-order sliding mode control (e.g., \cite{Basin-2016, A.Chalanga_S.Kamal_L.M.Fridman_B.Bandyopadhyay_J.A.Moreno}). The key to guarantee prescribed-time consensus in our approach is to enforce the tracking error to be null, despite the disturbances, before the preset settling time $t_f$. This can be achieved by using a robust fixed-time control for high-order systems instead of the term $\nu_i$ in \eqref{stc_control}, for instance, by using the discontinuous controller of \cite{Mishra-2018a} or the continuous scheme of \cite{Mishra-2018b}. Let us illustrate the application of the controller of \cite{Mishra-2018b} in our prescribed-time consensus approach. 

Consider the controller $v_i$ and the term $\mu_i$ of Theorem~\ref{theo_stc_high_order}, but $\nu_i$ is now defined according to the following expressions from \cite{Mishra-2018b}:

The sliding surface is now defined by the parameters $c_j, b_j, \alpha_j$ and $\beta_j$ for $j=1,...,n$ as
\begin{align}
    s_i&=\dot{\xi}_{i_n}+c_n|\xi_{i_n}|^{\alpha_n}\sign(\xi_{i_n})+c_{n-1}|\xi_{i_{n-1}}|^{\alpha_{n-1}}\sign(\xi_{i_{n-1}}) \nonumber \\
    &+...+c_1|\xi_{i_1}|^{\alpha_1}\sign(\xi_{i_1})+b_n|\xi_{i_n}|^{\beta_n}\sign(\xi_{i_n}) \nonumber \\
    &+b_{n-1}|\xi_{i_{n-1}}|^{\beta_{n-1}}\sign(\xi_{i_{n-1}})+...+b_1|\xi_{i_1}|^{\beta_1}\sign(\xi_{i_1}).
\end{align}

The auxiliary control law is given by
\begin{align}
    \nu_i=u_{i_{eq}}+u_{i_n},
    \label{eq:fixed-timeAuxCont} 
\end{align}
with
\begin{align}
    u_{i_{eq}}&=c_n|\xi_{i_n}|^{\alpha_n}\sign(\xi_{i_n})+c_{n-1}|\xi_{i_{n-1}}|^{\alpha_{n-1}}\sign(\xi_{i_{n-1}}) \nonumber \\
    &+...+c_1|\xi_{i_1}|^{\alpha_1}\sign(\xi_{i_1})+b_n|\xi_{i_n}|^{\beta_n}\sign(\xi_{i_n}) \nonumber \\
    &+b_{n-1}|\xi_{i_{n-1}}|^{\beta_{n-1}}\sign(\xi_{i_{n-1}})+...+b_1|\xi_{i_1}|^{\beta_1}\sign(\xi_{i_1}),
\end{align}
and
\begin{align}
    \dot{u}_{i_n}=K\sign(s_i),  
\end{align}
this controller ensures fixed-time convergence to $s_i=0$ provided $K>\beta_i|\dot{\rho}_i|-\sum_{j \in N_i\setminus\{l\}}a_{ij} |\dot{\rho}_j|$ and some conditions on the parameters $c_j, b_j, \alpha_j, \beta_j$  $\forall j=1,...,n$ (see \cite{Mishra-2018b}). Furthermore, this implies convergence of the tracking error of each agent to the origin $\xi_i=\bld{0}_n$ in fixed-time with a known convergence bound $T_{max}$. If $T_{max}<t_f$, prescribed time-consensus will be achieved with \emph{continuous control signals}.

\remark{As reviewed in the introduction, to the best of our knowledge, only the works~\cite{Y.Zhao_Y.Liu_G.Wein_W.Ren_G.Chen,Y.Zhao_Y.Liu_a,Y.Wang_Y.Song, Lu-2020} have investigated prescribed-time consensus for high-order MASs. The proposed consensus protocol \eqref{stc_control} has the advantage, over those protocols, of providing robustness against large matched perturbations and by using the auxiliary control law \eqref{eq:fixed-timeAuxCont}, the whole control effort is continuous. Moreover, the protocols proposed in this work do not require information about the network's connectivity, as required by the referred protocols of the literature.}

\remark{It is worth noting that some aspects have to be considered during the application of the proposed consensus protocols. First, all the clocks of the agents in the network must be synchronized to achieve prescribed-time convergence. Second, physical constraints of the systems must be taken into account to set $t_f$, considering that a small $t_f$ will result in large control efforts. Thus, the maximum allowable input of each agent must be taken into consideration to set $t_f$, however, this is not in the scope of this work. Further analysis is required to obtain a relation between the maximum control effort $\max(|\bld{v}|)$ as a function of the prescribed settling time $t_f$ and the initial consensus error, however, we already know that this relation is linear for unperturbed cases when the settling time $t_f$ is fixed. Third, it is assumed that the state of each agent is available and transmitted to its neighbors without time-delay or packet dropouts.}

\remark{
\label{rem:undirectedGraphsLoopProblem}
The evaluation of the proposed control laws requires the knowledge of the control actions of neighboring agents. This can be achieved, from the theoretical point of view, by considering only the spanning tree in the communication graph, thus an agent only requires information from its child agents, avoiding thus communication loops. However, in practical implementations, this feature requires a very fast agents communication. Recent works have proposed to solve it using an observer as a first step of the control strategy, e.g. \cite{You-2020,Shi-2020}. A less costly and practical solution is the use of a communication buffer to allow each agent to use the inputs from the last time instant, as done in \cite{B.Tian_Z.Zuo_H.Wang, Ning-2020}. 
We will show in simulations that the last option makes possible the application of our approach for both directed graphs and connected \emph{undirected graphs} (where communication loops exist), however, further analysis must be done to formally solve the communication loop problem.} 


\section{Simulations}\label{sec:simulations}

In this section, simulations are performed to illustrate the effectiveness and advantages of the proposed TBG-based consensus protocols. The simulations
were implemented in MATLAB using the Euler forward
method to approximate the time derivatives with a time step of $0.1$ms. No special functions of this software were used. We present results using the nonlinear controller \eqref{stc_control}, which enhances the performance of the linear controller \eqref{eq:controller_lin} due to its robustness properties. In the following simulations, a MAS of 8 third-order agents is considered, where the agents' dynamics are described by \eqref{chainInt_nonlinear} with $f(\bld{x}_i) = x_{i1}x_{i2}\sin(x_{i3}) + 0.1x_{i1}x_{i3}$ and $g(\bld{x}_i) = -2$ for agents $\{1,3,5,7\}$, and $f(\bld{x}_i) = 0$ and $g(\bld{x}_i) = 1$ for agents $\{2,4,6,8\}$. None of the agents exhibit zero dynamics (i.e., the variables $\bld{\varphi}_i$ in \eqref{chainInt_nonlinear} do not exist). The convergence time is preset to $t_f = 5$ seconds. The implementation of the proposed TBG-tracking protocols requires the computation of the TBGs references and the time-dependent gain, i.e., to design the functions $h_1(t),\dots,h_n(t)$ fulfilling \eqref{hfunction}, and to evaluate $\bld{H}(t)$ and $\bld{K}_t(t)$ during the system evolution. The following TBG functions are used, $h_1(t) = 20(t/t_f)^7 -70(t/t_f)^6 + 84(t/t_f)^5 -35(t/t_f)^4 +1$, $h_2(t) = 10t^7/t_f^6 -36t^6/t_f^5 + 45t^5/t_f^4 -20t^4/t_f^3 + t$, $h_3(t)= 2t^7/t_f^5 -7.5t^6/t_f^4 + 10t^5/t_f^3 -5t^4/t_f^2 + t^2/2$, which fulfill with \eqref{hfunction}.

\begin{figure}[ht]
\begin{center}
\includegraphics[width=3.5cm]{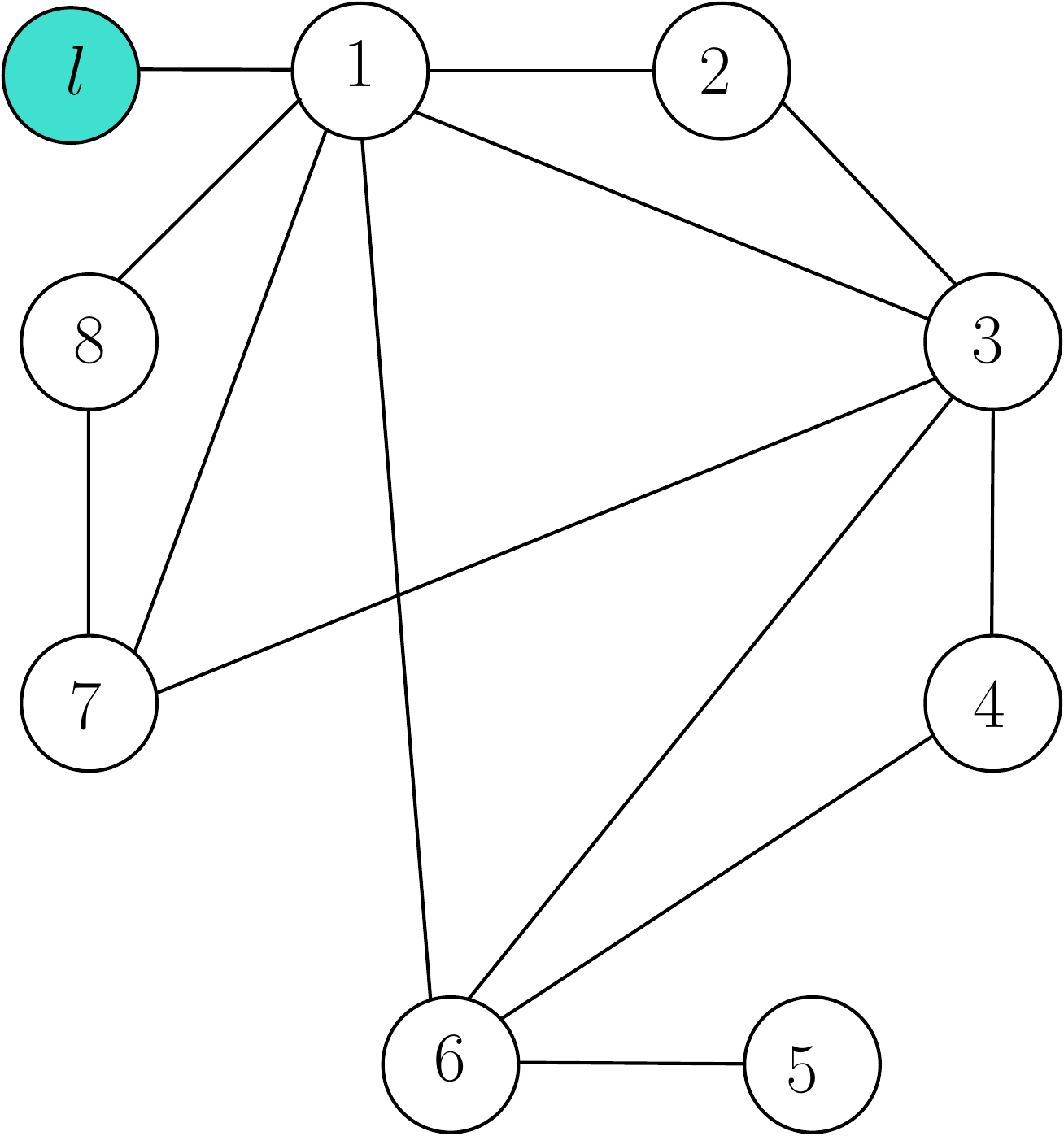} \hspace{0.5cm}
\includegraphics[width=3.5cm]{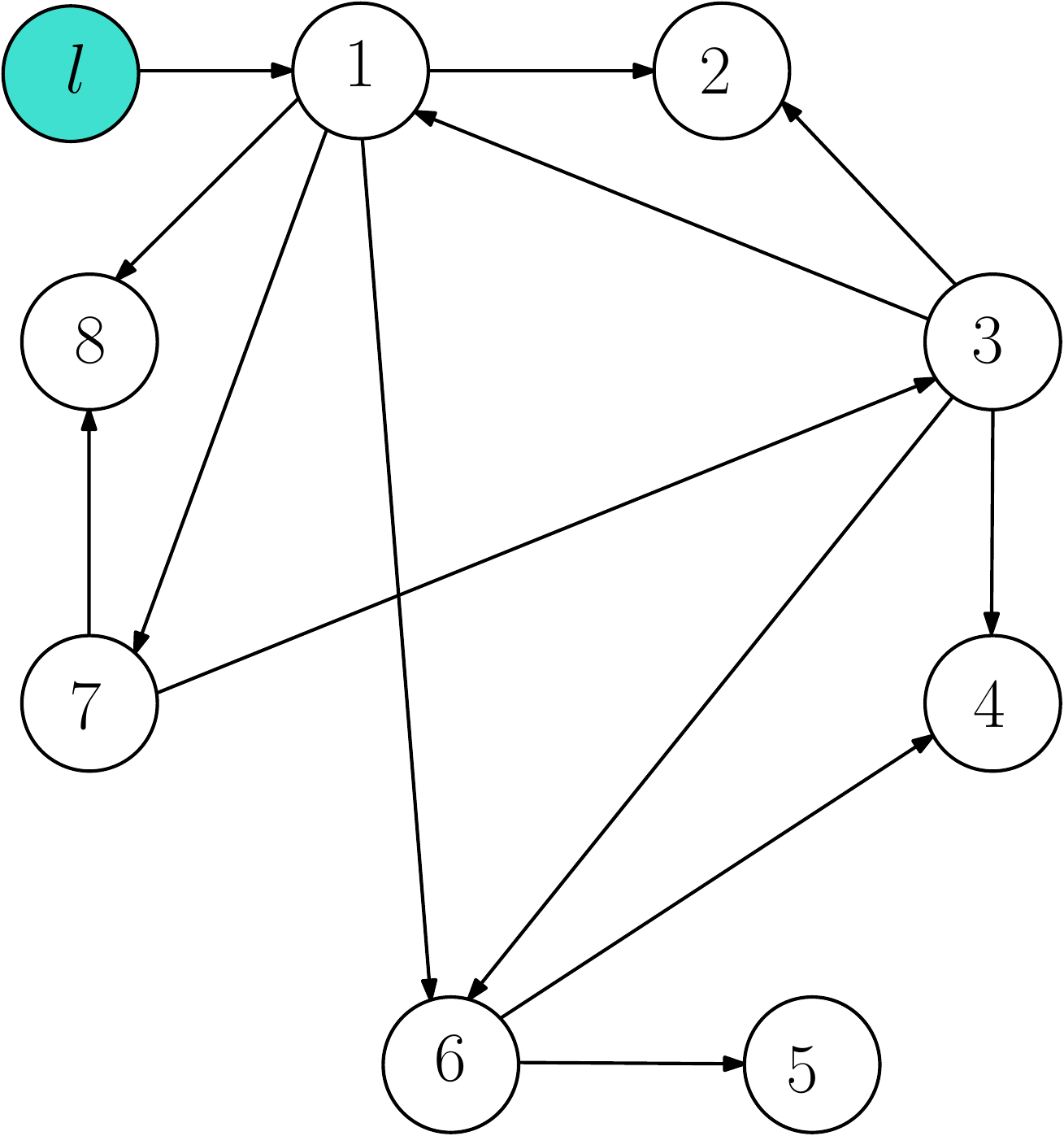}
\caption{Communication graphs taken from \cite{C.Yong_X.Guangming_L.Huiyang}. Left: undirected graph $\mathcal{G}_1$. Right: directed graph $\mathcal{G}_2$. }
\label{fig_graphs}
\end{center}
\end{figure}

The communication topologies shown in fig. \ref{fig_graphs} will be used~(\cite{C.Yong_X.Guangming_L.Huiyang}). The first one $\mathcal{G}_1$ is a connected undirected graph, the second one $\mathcal{G}_2$ is a directed graph having a spanning tree. Both topologies consist of 1 leader and 8 followers. As described in Remark \ref{rem:undirectedGraphsLoopProblem}, both kinds of graphs allow us to solve the consensus tracking problem in prescribed time; however, the undirected graph considers more communication requirements due to its bidirectional connections. On the one hand, as proved in our theorems, it is sufficient and necessary that the graph has a directed spanning tree, with the advantage of avoiding the communication loop problem, as detailed in the referred remark above. On the other hand, an undirected graph has the advantage that its extra connections, with respect to a directed graph with the same vertices, may make the MAS more robust against communication failures. The last requires further analysis as a consensus problem with switching topologies.

The initial states of the eight agents are randomly selected in the range $(-2,2)$ and are shown in Table \ref{x0_high_order_table}. To avoid communication loops, control laws are evaluated by using the information of neighbor agents available from the previous time instant.

\begin{table}
\caption{Initial states of the 8 agents with third-order dynamics.}
\begin{center}
\scriptsize{
\begin{tabular}{*{9}{|c}|}
\hline
$i$ & \multicolumn{1}{c|}{1} & \multicolumn{1}{c|}{2} & \multicolumn{1}{c|}{3} & \multicolumn{1}{c|}{4} & \multicolumn{1}{c|}{5} & \multicolumn{1}{c|}{6} & \multicolumn{1}{c|}{7} & \multicolumn{1}{c|}{8} \\ \hline
$x_{i1}(0)$ & -1.66 & 1.89 & 0.60  & -1.07  & -0.38 & -1.51  & -0.92 & -0.96  \\ 
$x_{i2}(0)$ & -0.67 & -1.39 & -0.60  & -1.51 & 1.53 & -1.62 & 1.72 & -0.40  \\ 
$x_{i3}(0)$ & -1.81  & -0.63 &  0.94  & 1.17 &  0.17 & 0.74 & 1.57 & -1.78 \\ \hline
\end{tabular}}
\end{center}
\label{x0_high_order_table}
\end{table}

\subsection{Proposed Prescribed-Time Consensus}

Since the results are similar for the unperturbed case, in this subsection, we show the performance of the proposed robust prescribed-time TBG controller \eqref{stc_control} considering disturbances $\rho_i(t)=\alpha_i(1 + 1\sin(5t))$, with $\alpha_i$ randomly selected in $(0,1)$. We evaluated the consensus protocol for both communication topologies $\mathcal{G}_1$ and $\mathcal{G}_2$. A third-order leader (root node) is considered, having communication only with the first follower agent, i.e., $b_1 = 1$ and $b_i = 0, \, \forall i = \left\lbrace 2,\dots,8 \right\rbrace$. The leader's behavior is modeled as in \eqref{dyn_leader}, and its state is maintained constant and equal to $\bld{x}_l(t) = [-1,0,0]^T$ by setting its control input as $u_l = 0$. The gain of the robust controller is set as $k_1 = 2.5$ and $\bld{K}_{fr} = [1, 2]$.

\begin{figure}[ht]
\begin{centering}
\includegraphics[width=8.0cm]{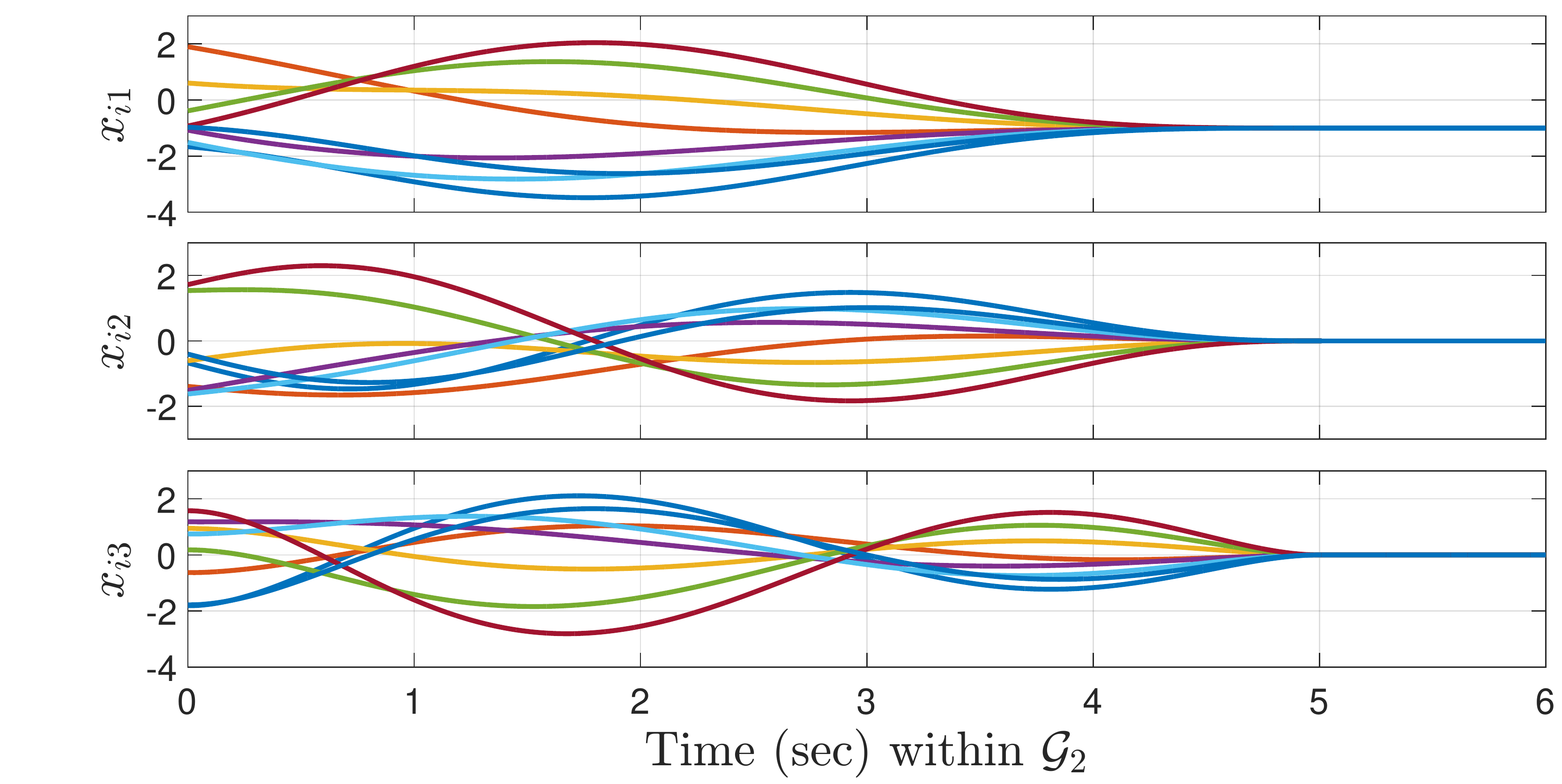} 
\includegraphics[width=8.0cm]{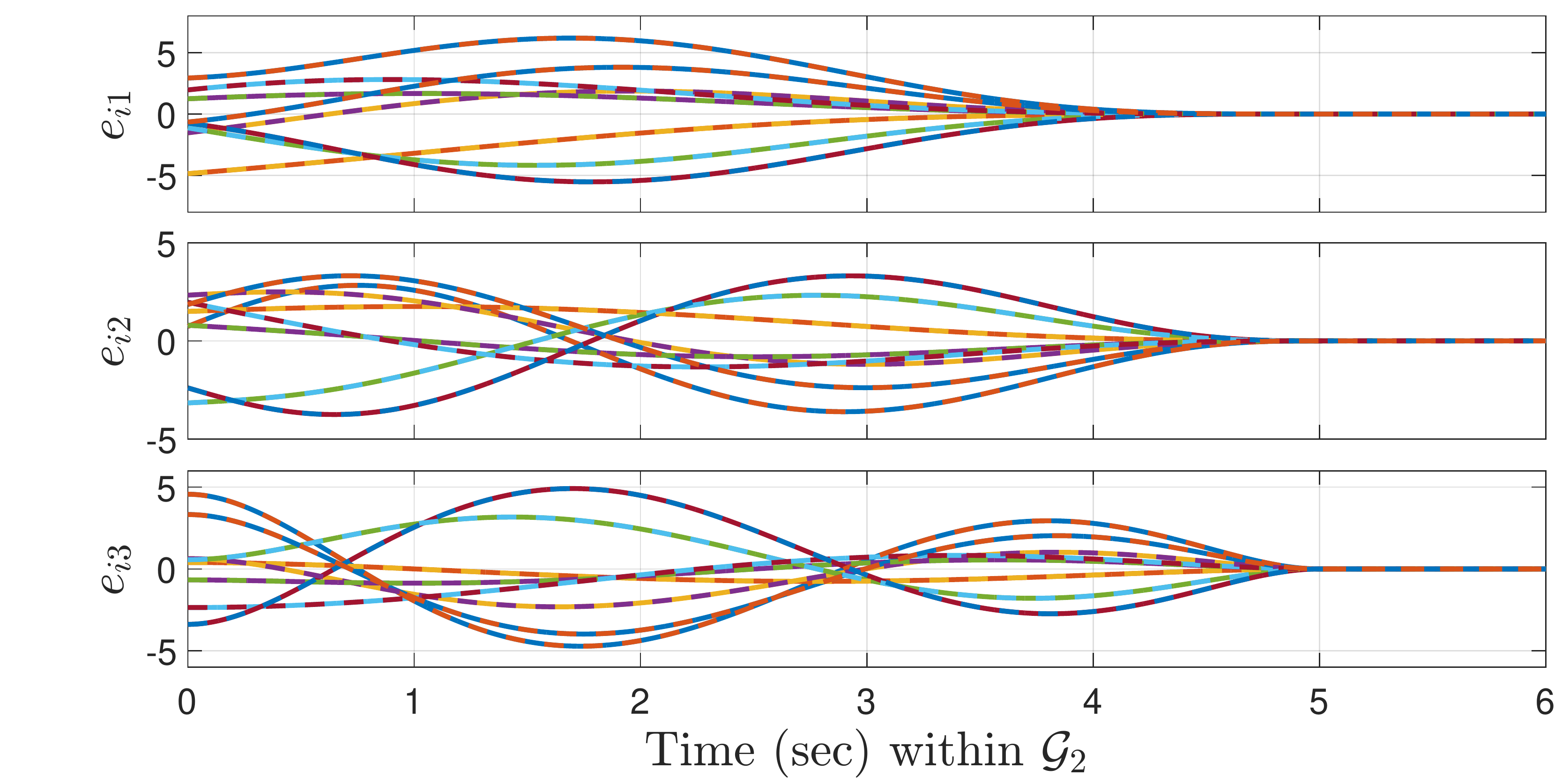} 
\caption{State evolution (top) and consensus error trajectories (bottom) of the third-order perturbed MAS under~\eqref{stc_control} with $\nu_i$ of \eqref{eq:fixed-timeAuxCont}, for $\mathcal{G}_2$ and $t_f = 5$s. The consensus state at $t_f$ is $\bld{x}_i(5) = \bld{x}_l(5) = [-1,0,0]^T$. In the bottom figure, the continuous lines represent the evolution of the errors whereas the TBG references are drawn with dashed lines.
}
\label{fig_stc_digraph_state_error}
\end{centering}
\end{figure}

\begin{figure}[ht]
\begin{centering}
\includegraphics[width=8.0cm]{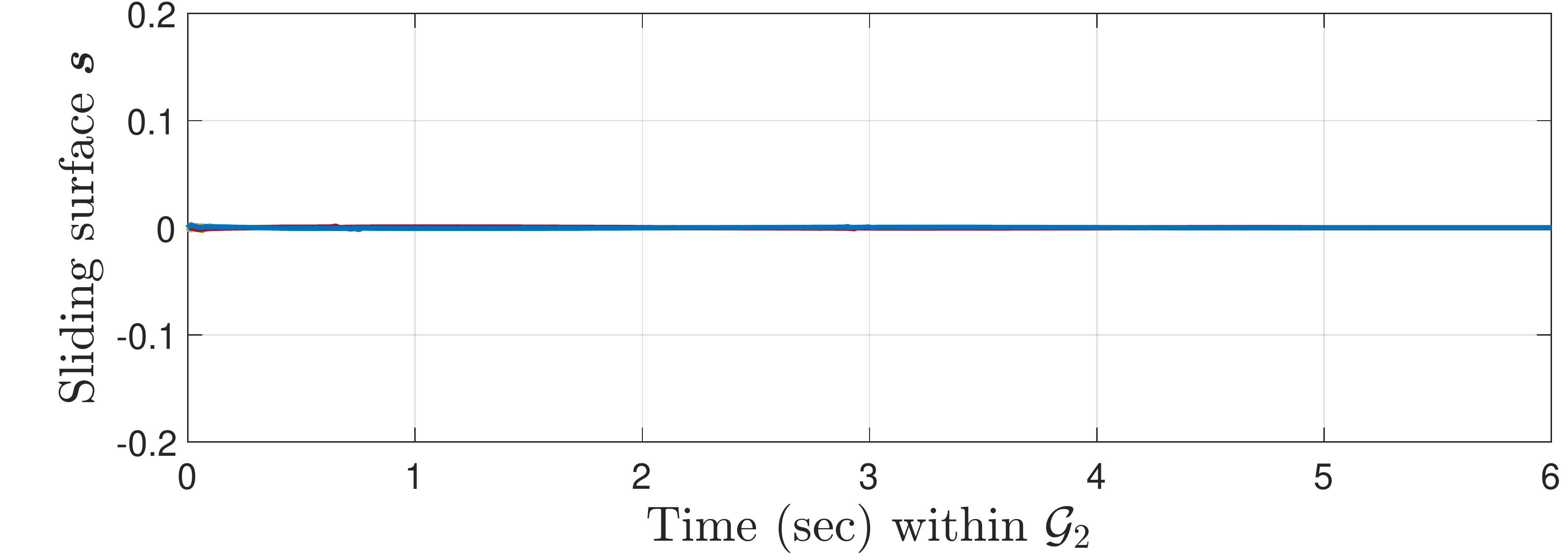} 
\includegraphics[width=8.0cm]{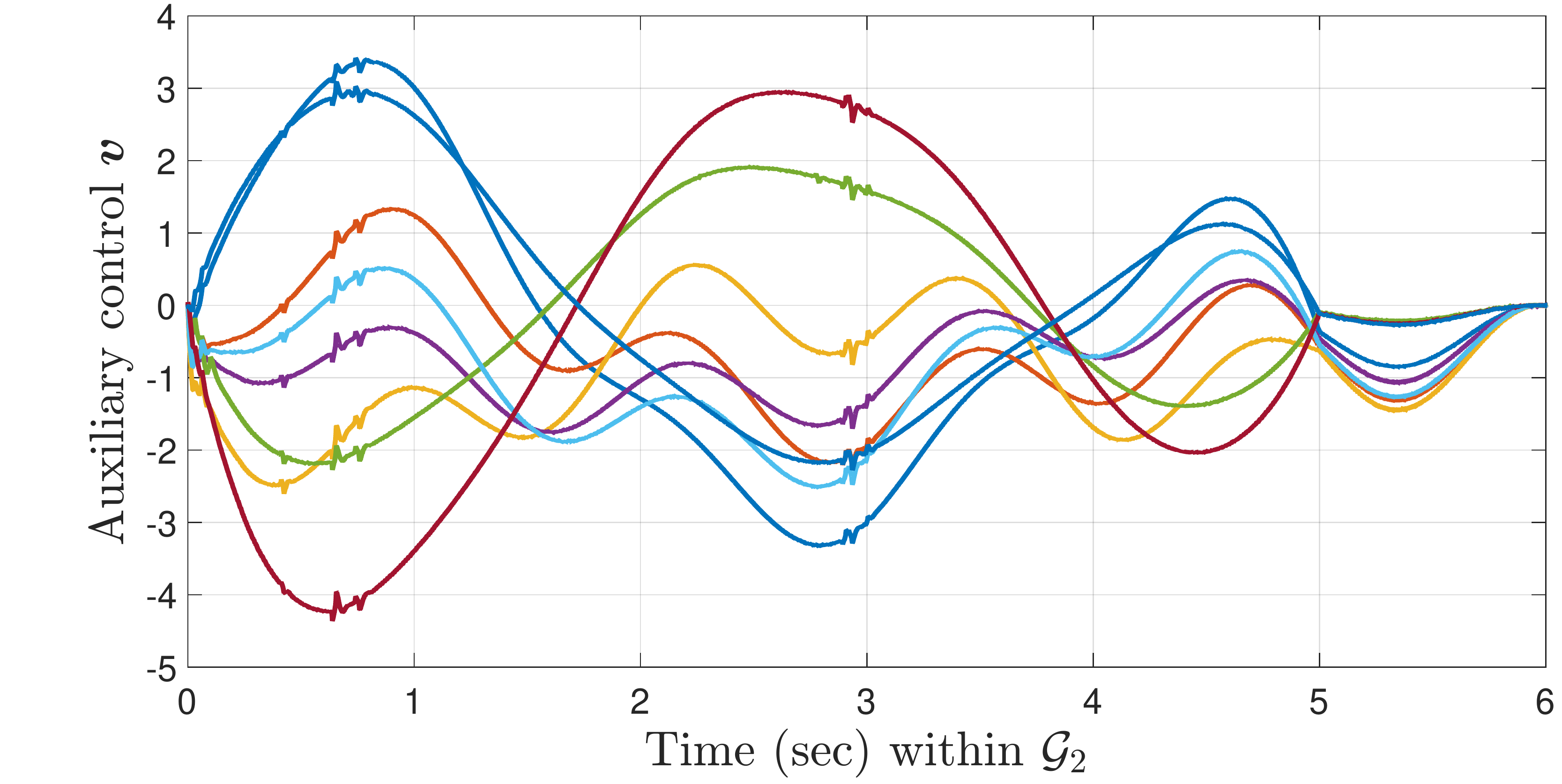} 
\caption{Sliding surfaces evolution (top) and auxiliary control inputs response (bottom) of the third-order perturbed MAS under~\eqref{stc_control} with $\nu_i$ of \eqref{eq:fixed-timeAuxCont}, for $\mathcal{G}_2$ and $t_f = 5$s.
}
\label{fig_stc_digraph_surface_input}
\end{centering}
\end{figure}

Figs. \ref{fig_stc_digraph_state_error}-\ref{fig_stc_digraph_surface_input} present the results for the directed communication graph with the described time-varying disturbance $\bld{\rho}(t)$. We introduced as the robustness term $\nu_i$ the continuous fixed-time controller \eqref{eq:fixed-timeAuxCont} to obtain a completely continuous prescribed-time control law. In particular, for the same third order MAS under evaluation, we set:
\begin{align}
    s_i=\dot{\xi}_{i_3}&+15|\xi_{i_3}|^{7/10}\sign(\xi_{i_3})+66|\xi_{i_2}|^{7/13}\sign(\xi_{i_2}) \nonumber \\
    &+80|\xi_{i_1}|^{7/16}\sign(\xi_{i_1})+15|\xi_{i_3}|^{21/20}\sign(\xi_{i_3}) \nonumber \\
    &+66|\xi_{i_2}|^{21/19}\sign(\xi_{i_2})+80|\xi_{i_1}|^{21/18}\sign(\xi_{i_1})
    \label{eq:slidsurfFTC}
\end{align}
and
\begin{align}
    u_{i_{eq}}=&15|\xi_{i_3}|^{7/10}\sign(\xi_{i_3})+66|\xi_{i_2}|^{7/13}\sign(\xi_{i_2}) \nonumber \\
    &+80|\xi_{i_1}|^{7/16}\sign(\xi_{i_1})+15|\xi_{i_3}|^{21/20}\sign(\xi_{i_3}) \nonumber \\
    &+66|\xi_{i_2}|^{21/19}\sign(\xi_{i_2})+80|\xi_{i_1}|^{21/18}\sign(\xi_{i_1}),
\end{align}
and $\dot{u}_{i_n}=10\sign(s_i)$, which has a convergence bound of $T_{max}=3$s (see example in \cite{Mishra-2018b}) that is enough for our aim of achieving consensus in preset $t_f=5$s.

As shown in fig. \ref{fig_stc_digraph_state_error} (top), the followers achieve consensus to the leader's state $\bld{x}_l=[-1,0,0]^T$ at the prescribed time $5$ seconds. The consensus error trajectories shown in fig. \ref{fig_stc_digraph_state_error} (bottom) start over the TBG references and they converge to the origin in the prescribed time. As can be seen in fig. \ref{fig_stc_digraph_surface_input} (top), the sliding surfaces, defined as \eqref{eq:slidsurfFTC}, initiate on zero. The disturbances make the MAS to evolve slightly out of the surfaces, however the control enforces $s_i=0$ for $t\geq 3$. Fig. \ref{fig_stc_digraph_surface_input} (bottom) shows the auxiliary control inputs $v_i$ of each agent, which evolve smoothly over time and keep oscillating after $t_f$ in order to reject the disturbance.

\begin{figure}[ht]
\begin{centering}
\includegraphics[width=8.0cm]{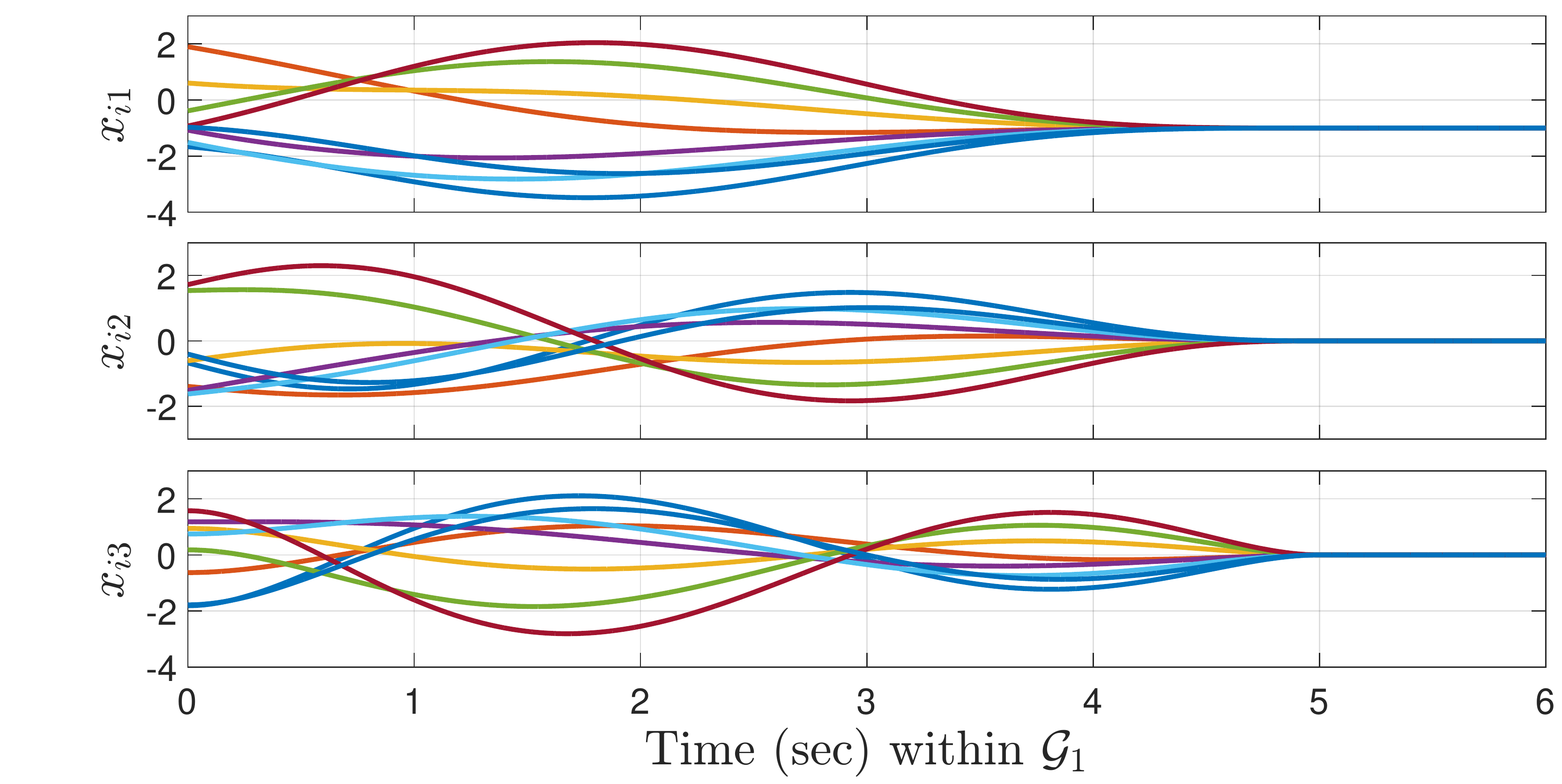} 
\includegraphics[width=8.0cm]{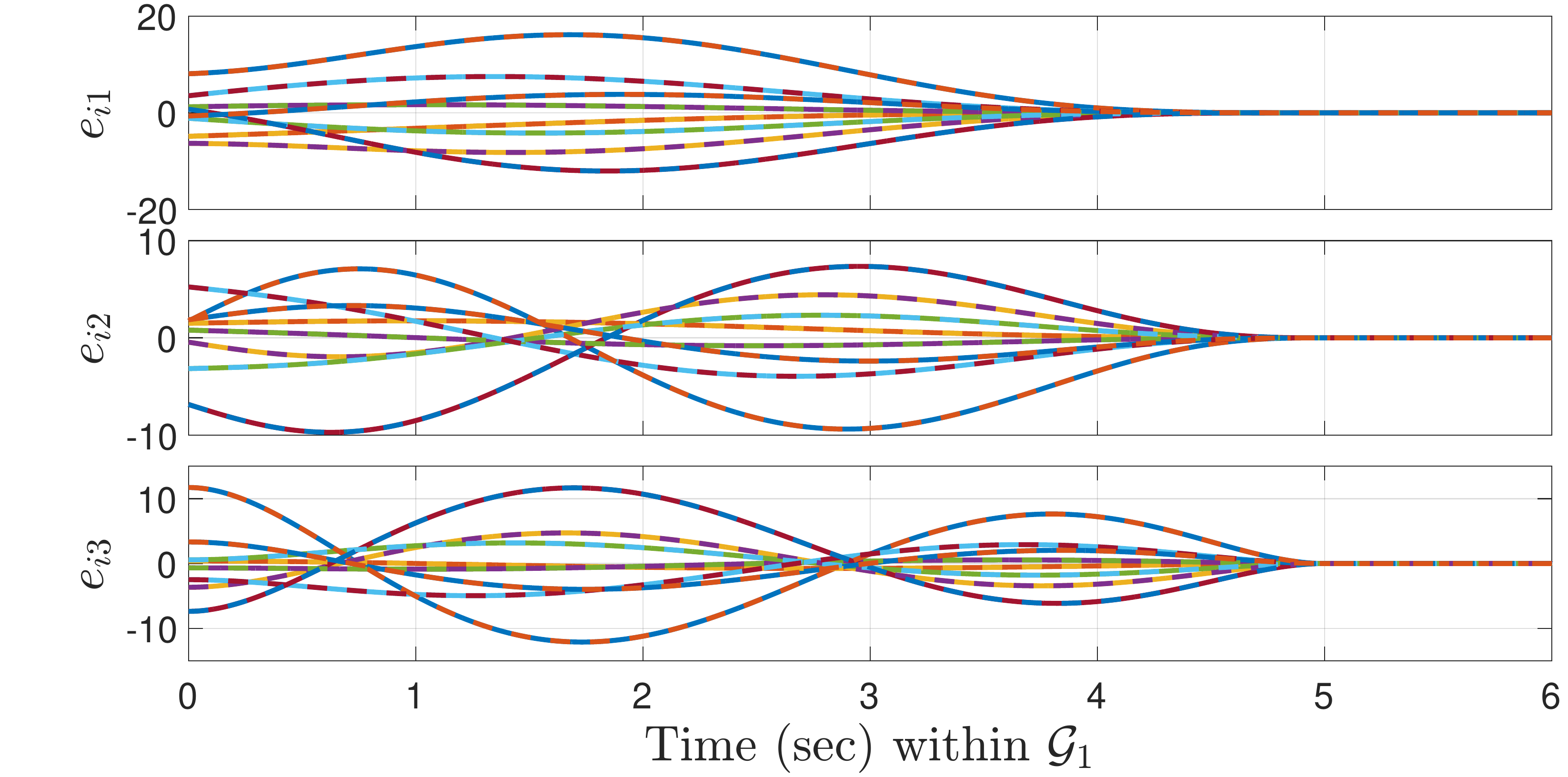}
\caption{State evolution (top) and consensus error trajectories (bottom) of the third-order perturbed MAS under~\eqref{stc_control}, for $\mathcal{G}_1$ and $t_f = 5$ s. The consensus state at $t_f$ is $\bld{x}_i(5) = \bld{x}_l(5) = [-1,0,0]^T$. In the bottom figure, the continuous lines represent the evolution of the errors whereas the TBG references are drawn with dashed lines.
}
\label{fig_stc_Undigraph_state_error}
\end{centering}
\end{figure}

\begin{figure}[ht]
\begin{centering}
\includegraphics[width=8.0cm]{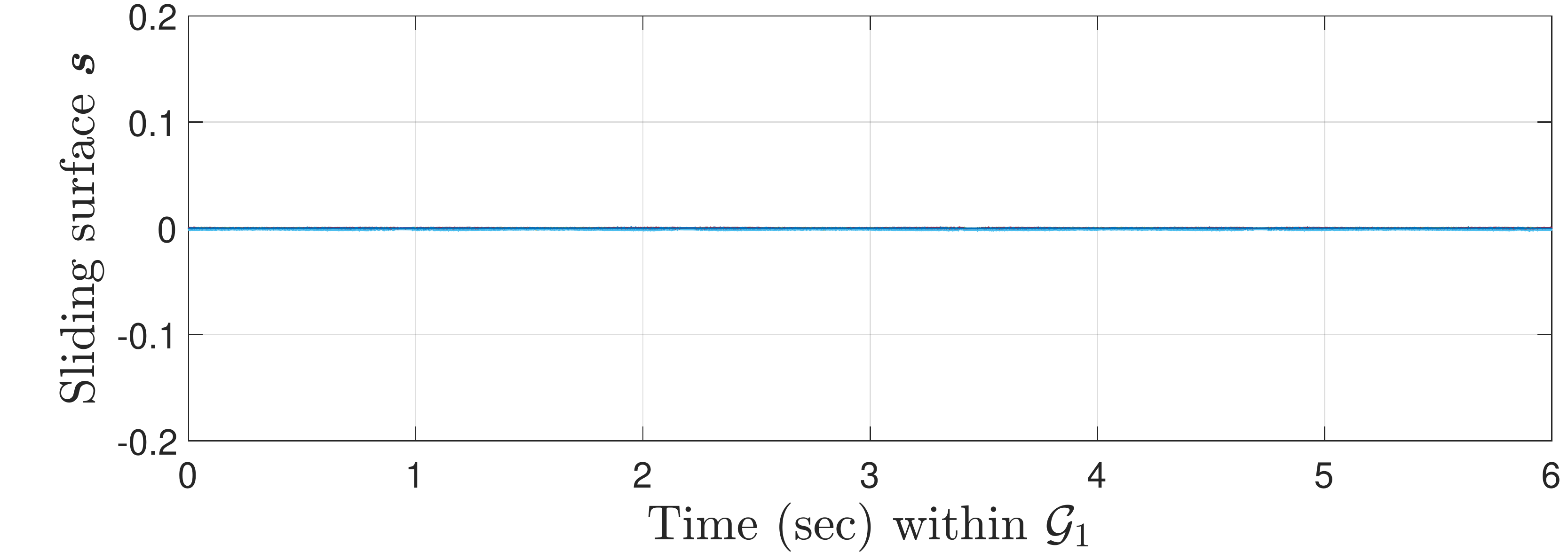} 
\includegraphics[width=8.0cm]{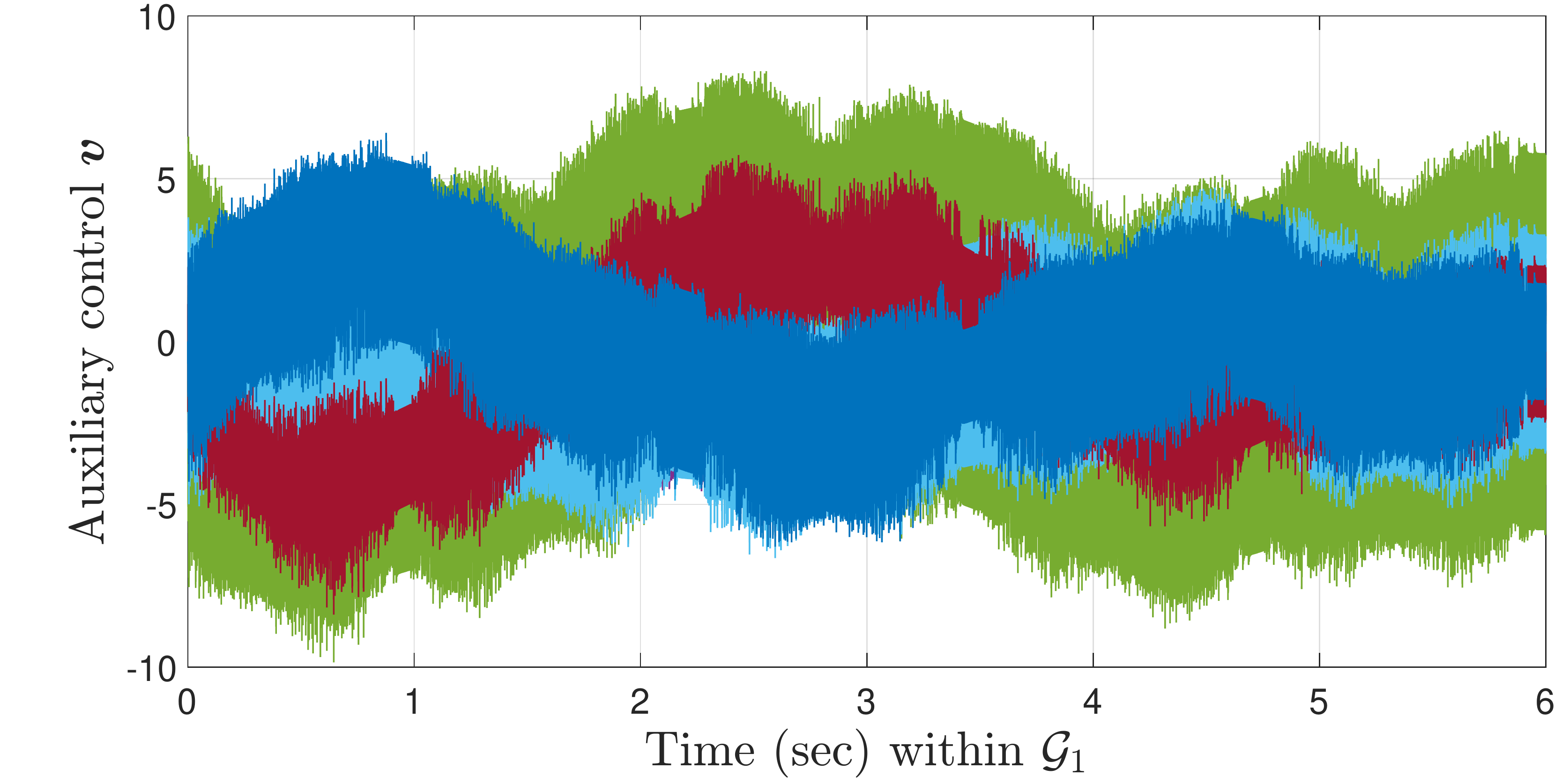}
\caption{Sliding surfaces evolution (top) and auxiliary control inputs response (bottom) of the third-order perturbed MAS under~\eqref{stc_control}, for $\mathcal{G}_1$ and $t_f = 5$s.
}
\label{fig_stc_Undigraph_surface_input}
\end{centering}
\end{figure}

Simulations results for the undirected communication graph are shown in figs. \ref{fig_stc_Undigraph_state_error}-\ref{fig_stc_Undigraph_surface_input}. The consensus performance under the robust TBG controller is shown in fig. \ref{fig_stc_Undigraph_state_error}. It can be observed in the top figure that, in spite of the disturbances $\bld{\rho}(t)$, the followers achieve consensus to the leader's state $\bld{x}_l=[-1,0,0]^T$ at the prescribed time of $5$ seconds. Fig. \ref{fig_stc_Undigraph_state_error} (bottom) shows that the consensus error trajectories track the TBG references, reaching the origin at the preset time. The sliding surfaces associated to each agent, computed as in \eqref{sliding_surface}, are shown in Fig. \ref{fig_stc_Undigraph_surface_input} (top). As explained in the Part II of the proof of Theorem \ref{theo_stc_high_order}, since the tracking errors initiate in zero and due to the stability of the sliding surface dynamics, the evolution of the tracking errors does not leave the manifold $s_i=0$. In this case, the robustness is achieved by the discontinuous term $\nu_i$ of \eqref{stc_control} as reflected in the auxiliary control inputs $v_i$ shown in Fig. \ref{fig_stc_Undigraph_surface_input} (bottom), which reject the disturbance vector $\bld{\rho}(t)$.

\subsection{Comparison with existing approaches}

\begin{figure}[ht]
\begin{centering}
\includegraphics[width=8.0cm]{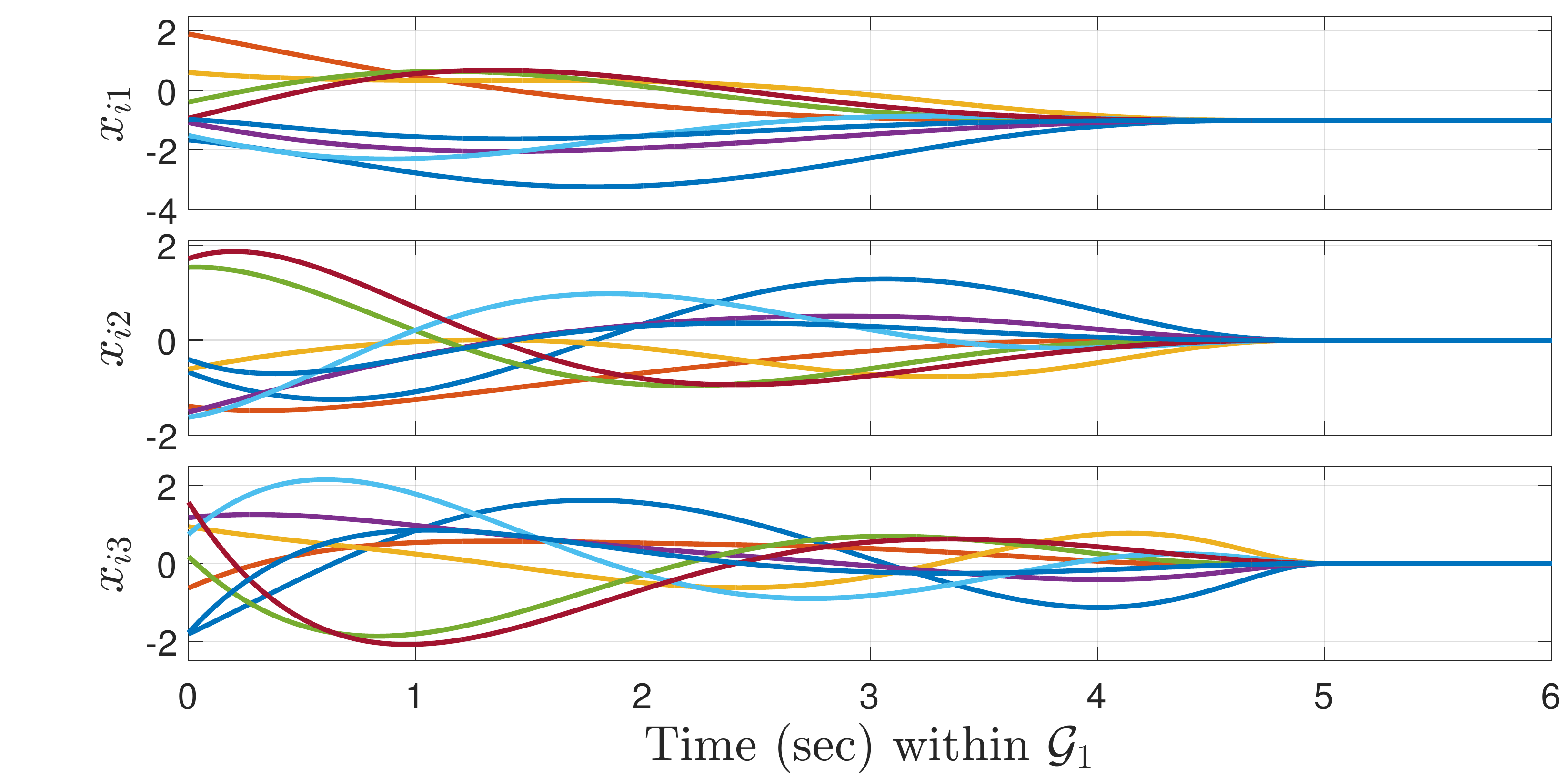} \includegraphics[width=8.0cm]{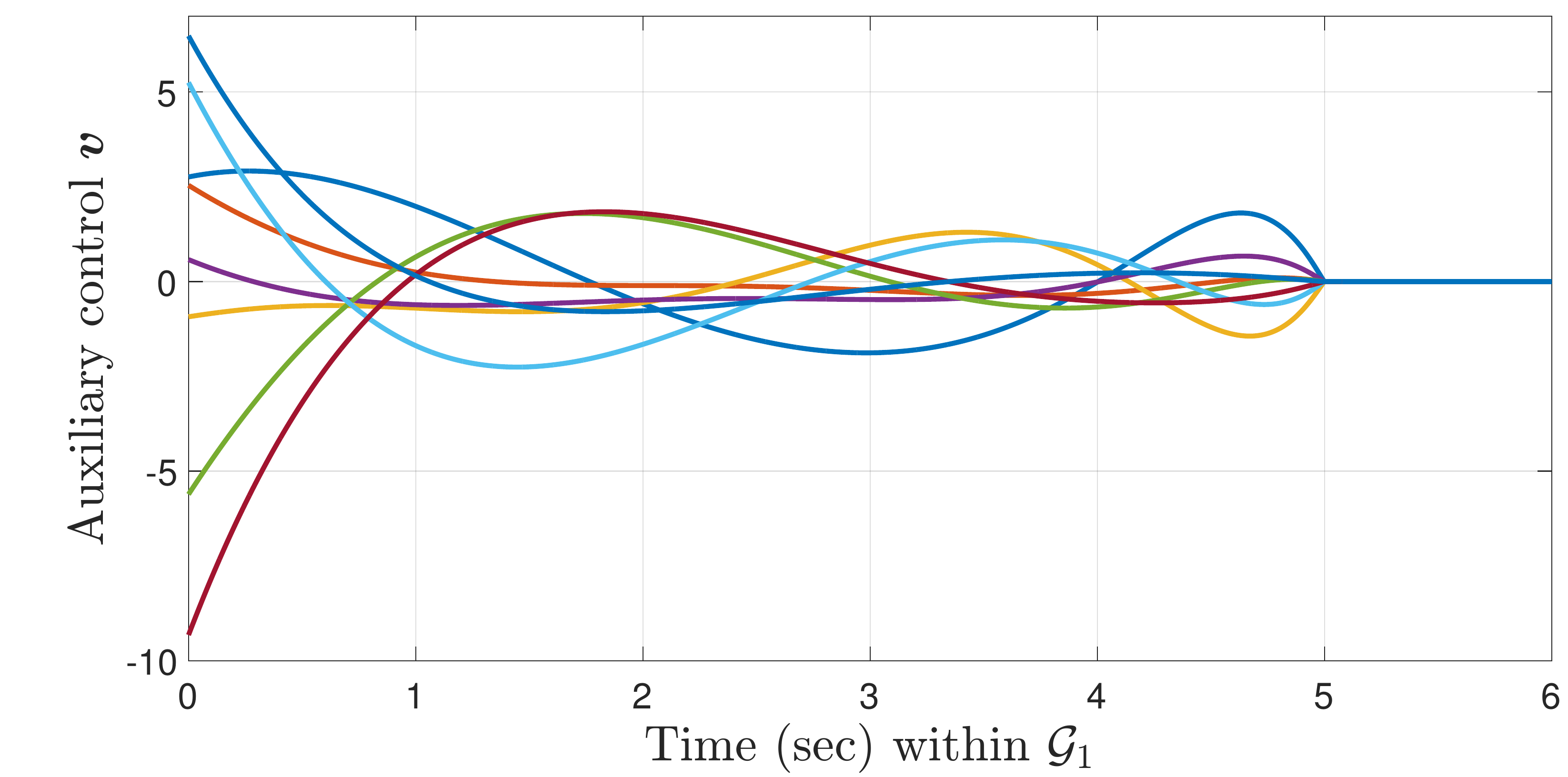} \caption{State response (top) and auxiliary control inputs (bottom) of the third-order MAS under the control of~\cite{Y.Wang_Y.Song} (eq. (19)), for $\mathcal{G}_1$ and $t_f = 5$s. The consensus state at $t_f$ is $\bld{x}_i(5) = \bld{x}_l(5) = [-1,0,0]^T$. 
}
\label{fig_Prespecified_leader_continuos_UndGraph}
\end{centering}
\end{figure}

\begin{figure}[ht]
\begin{centering}
\includegraphics[width=8.0cm]{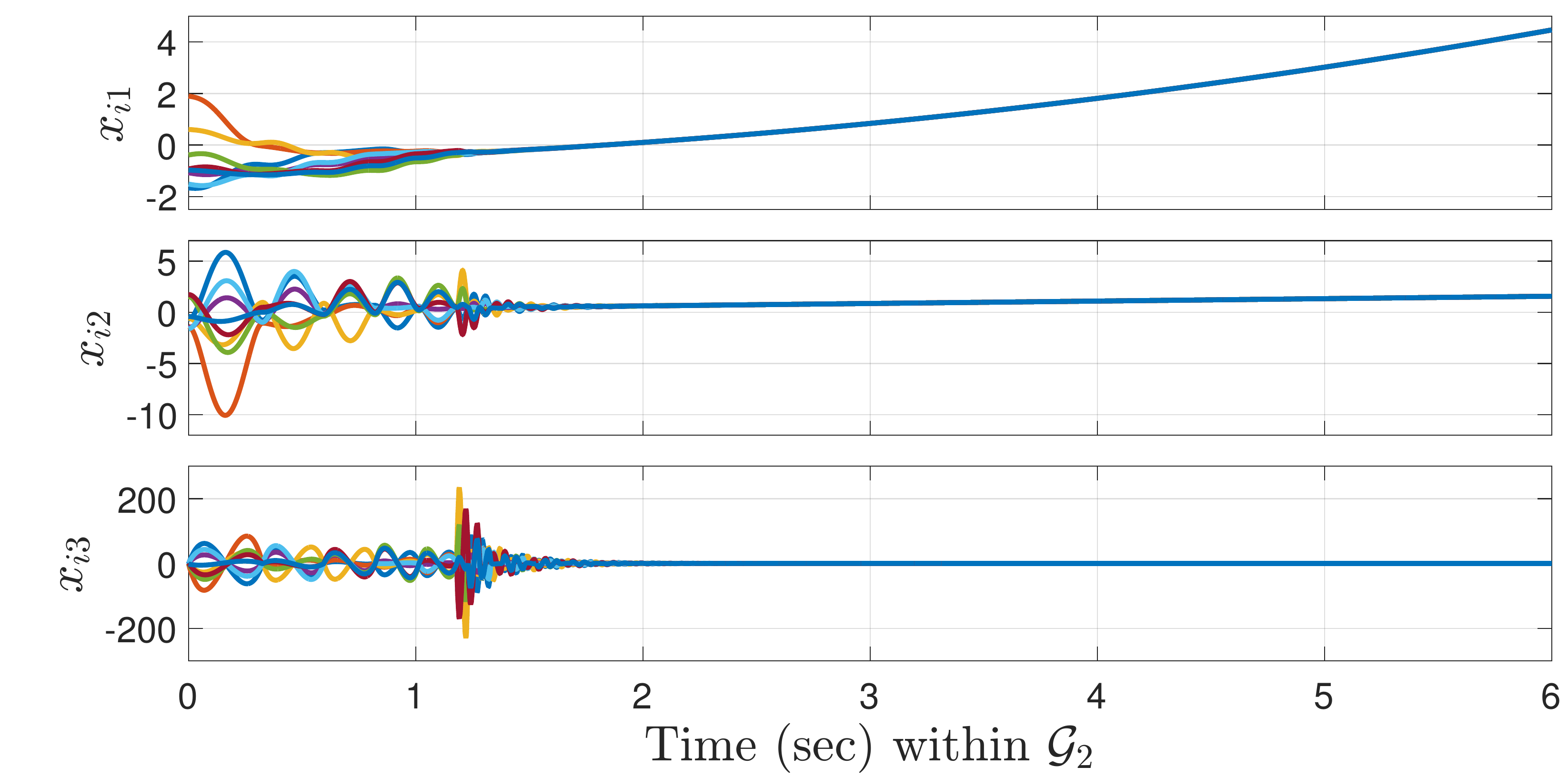} \includegraphics[width=8.0cm]{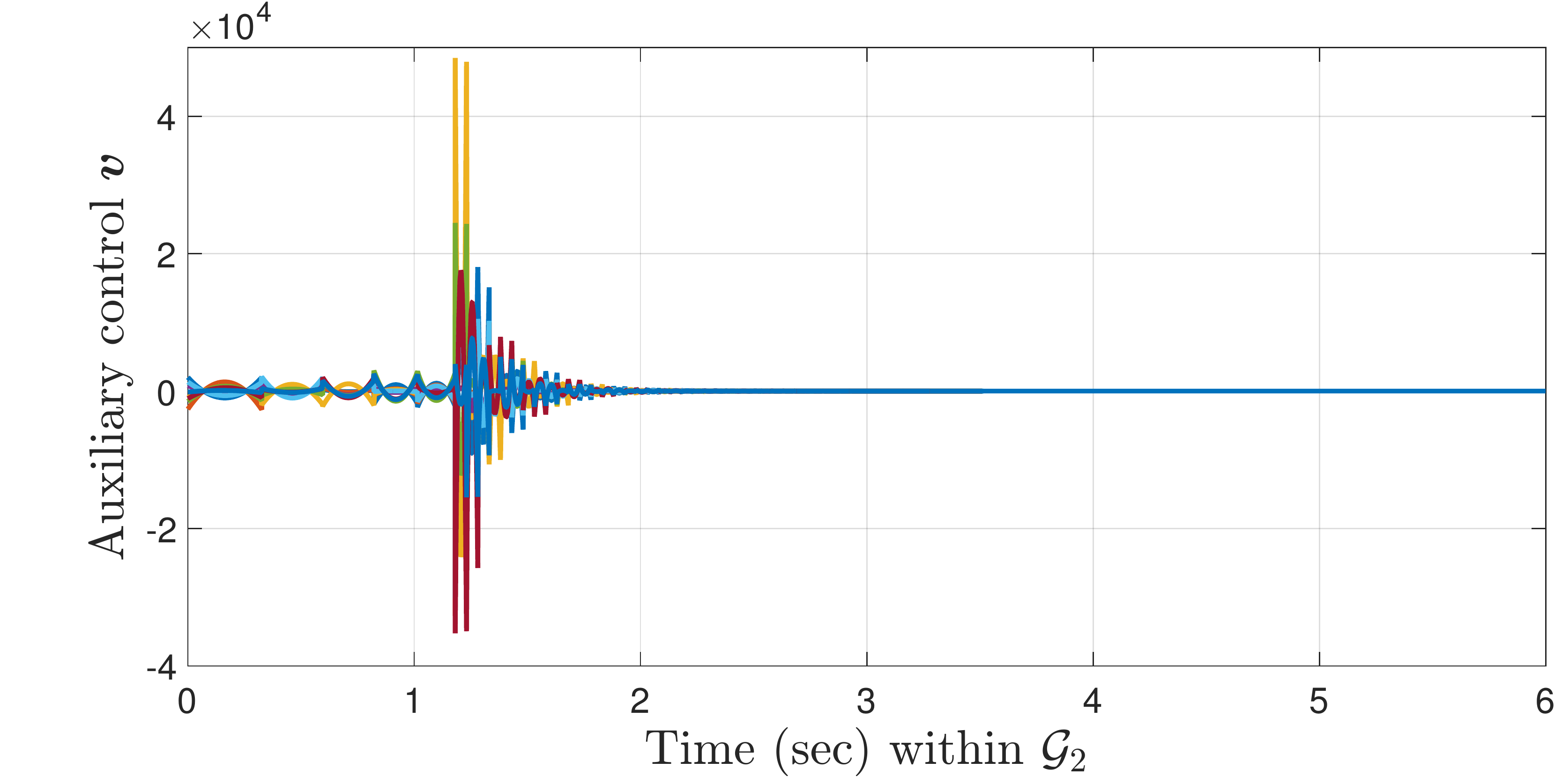} \caption{State response (top) and auxiliary control inputs (bottom) of the third-order MAS under the control of~\cite{Y.Zhao_Y.Liu_G.Wein_W.Ren_G.Chen} (eq. (2)), for $\mathcal{G}_2$ and $t_f = 5$s. The consensus state at $t_f$ is $\bld{x}_i(5) = \bld{x}^*_i = [ 3.01,1.32,0.23]^T$. 
}
\label{fig_Specified_DiGraph}
\end{centering}
\end{figure}

\begin{figure}[ht]
\begin{centering}
\includegraphics[width=8.0cm]{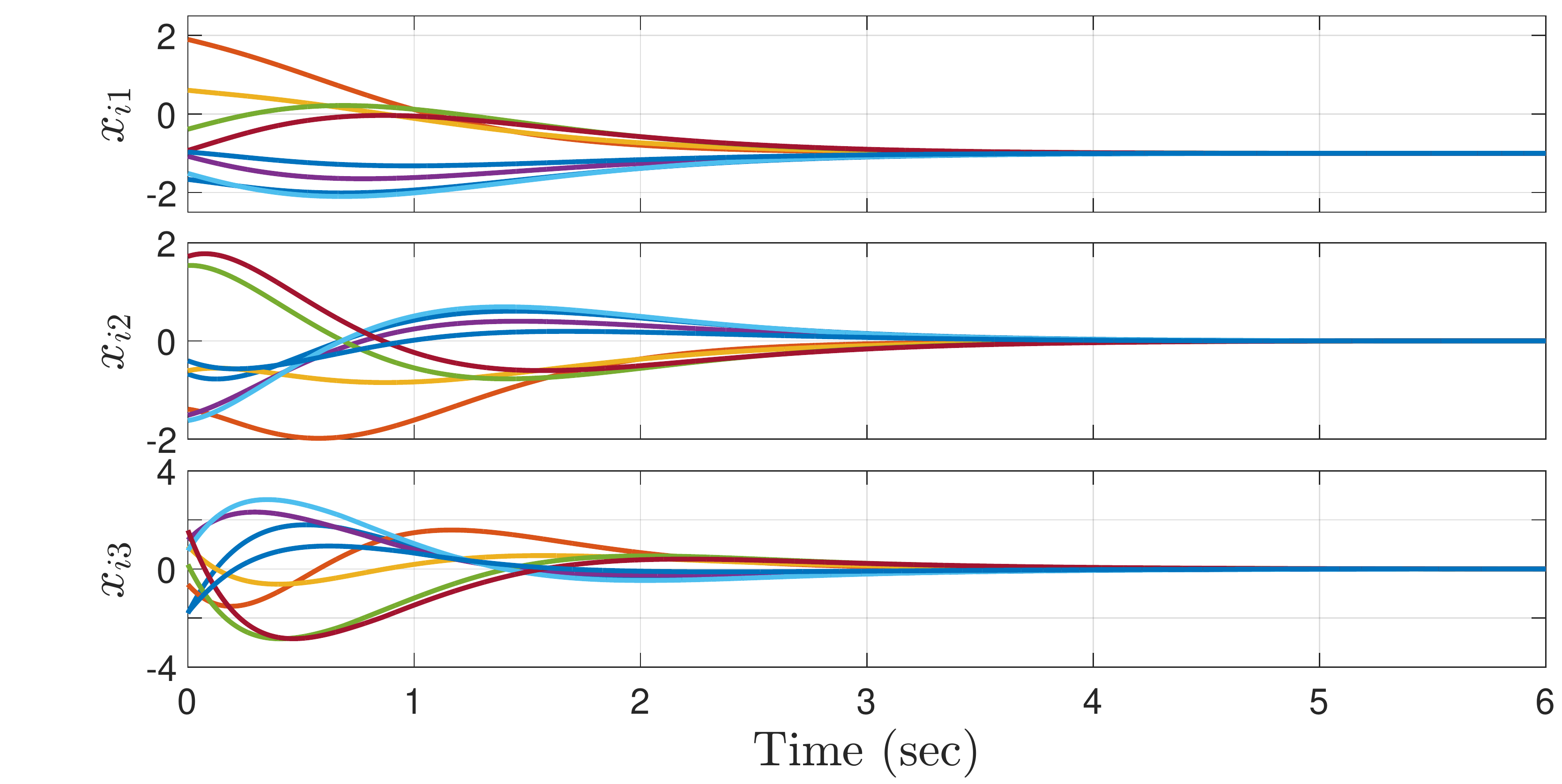} \includegraphics[width=8.0cm]{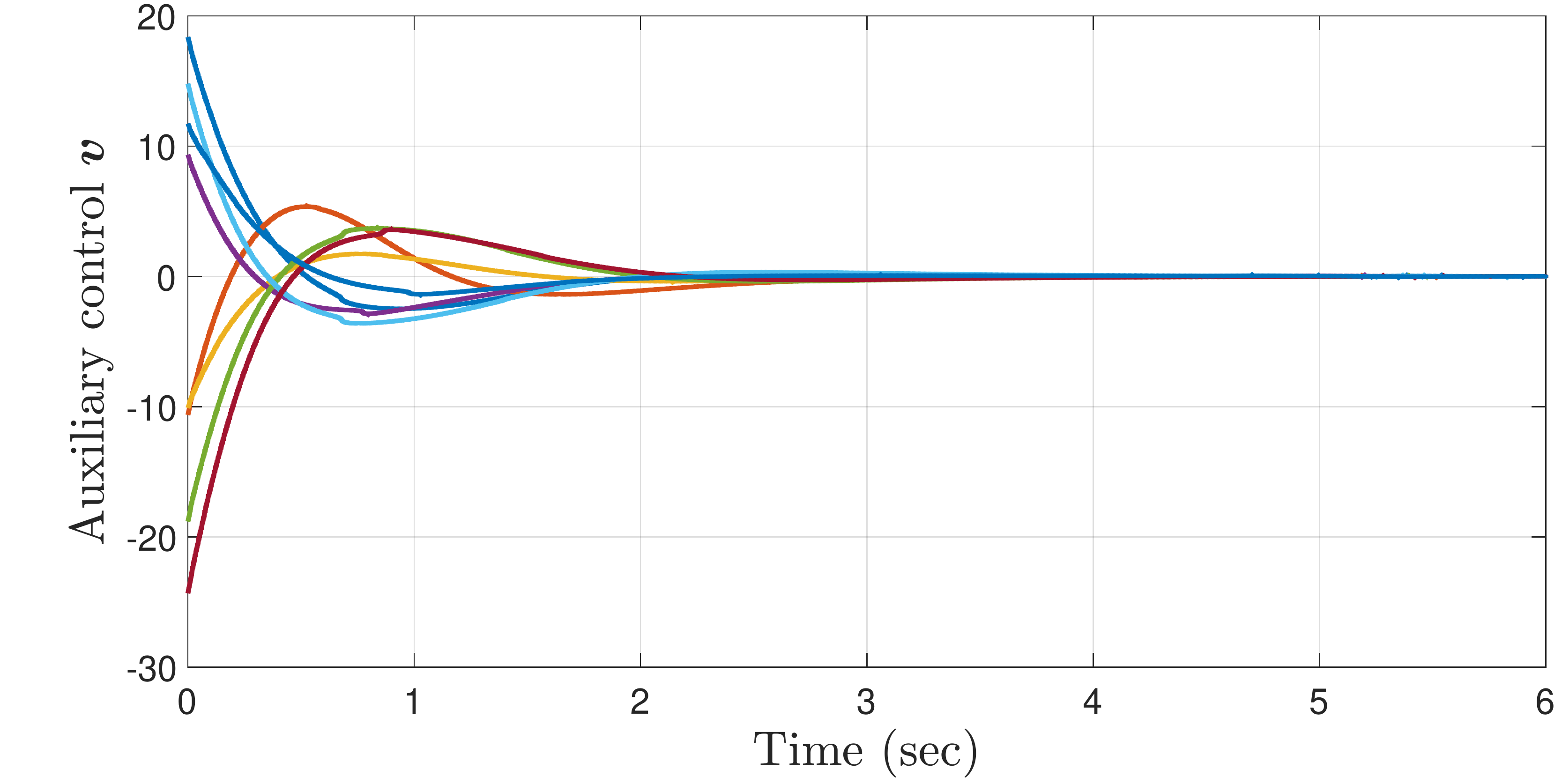} \caption{State response (top) and auxiliary control inputs (bottom) of the third-order systems under the control of~\cite{Z.Zuo_B.Tian_M.Defoort_Z.Ding} (eq. (30)) without the observer term. The final state at $t_f$ is $\bld{x}_i(5) = \bld{x}_l(5) = [-1,0,0]^T$. 
}
\label{fig_Fixed_Time}
\end{centering}
\end{figure}

For comparison purposes, simulations of the prescribed-time consensus protocols for high-order systems presented in \cite{Y.Zhao_Y.Liu_G.Wein_W.Ren_G.Chen,Y.Wang_Y.Song} are given in this subsection. In addition, simulations of a well-known fixed-time algorithm (\cite{Z.Zuo_B.Tian_M.Defoort_Z.Ding}) are included. The leader-following scheme in \cite{Y.Wang_Y.Song} is based on a time-varying scaling function, with a parameter to set, and a matrix defined by optimal control. The scheme in \cite{Y.Zhao_Y.Liu_G.Wein_W.Ren_G.Chen} is a leaderless discrete-time protocol based on the infinite frequency sampling of a time sequence. For a realistic implementation of the last scheme, we evaluate the sampling truncation based on a consensus error bound. The approach in~\cite{Z.Zuo_B.Tian_M.Defoort_Z.Ding} is a two step fixed-time consensus tracking control. First, the method estimates the leader's state using a distributed observer, then a fixed-time controller is introduced to drive all followers to the estimated leader's state at a settling time uniformly bounded by $T_{max}$, which is defined according to several control parameters. For comparison purposes, we implemented the fixed-time controller assuming that the leader’s state is available to the followers. We set its control parameters to get a settling time around $t_f=5$s for all the followers. To have a fair comparison, all the approaches were implemented without the disturbance $\bld{\rho}(t)$, because~\cite{Y.Zhao_Y.Liu_G.Wein_W.Ren_G.Chen} and~\cite{Y.Wang_Y.Song} are not able to deal with it, although our approach and the one in~\cite{Z.Zuo_B.Tian_M.Defoort_Z.Ding} are capable to reject lumped uncertainties. The results are shown in figs.~\ref{fig_Prespecified_leader_continuos_UndGraph}-\ref{fig_Fixed_Time}, where the plots at the top of both figures show the state response and the plots at the bottom present the auxiliary control signals produced by the corresponding evaluated protocol. It can be observed in the top of figs.~\ref{fig_Prespecified_leader_continuos_UndGraph}-\ref{fig_Fixed_Time} that the states of all the agents converge at the settling time $t_f = 5$s to the consensus states $\bld{x}_l = [-1,0,0]^T$ for the approaches in \cite{Y.Wang_Y.Song} and \cite{Z.Zuo_B.Tian_M.Defoort_Z.Ding}, and $\bld{x}_i^* = [3.01,1.32,0.23]^T$ for the approach in \cite{Y.Zhao_Y.Liu_G.Wein_W.Ren_G.Chen}. 
Notice at the bottom of figs.~\ref{fig_Prespecified_leader_continuos_UndGraph}-\ref{fig_Fixed_Time} that the magnitude of the auxiliary control efforts for the compared control schemes are initially large, and particularly, the control efforts with the protocol of \cite{Y.Zhao_Y.Liu_G.Wein_W.Ren_G.Chen} become very large and the signals are not smooth due to its motion planning switching strategy. In comparison, fig. \ref{fig_stc_digraph_surface_input} shows that our proposed control protocol generates control signals that start in zero, provide a smooth continuous evolution and exhibit lower magnitudes compared with the existing prescribed-time consensus approaches. 

\begin{figure}[ht]
\begin{centering}
\includegraphics[width=8.0cm]{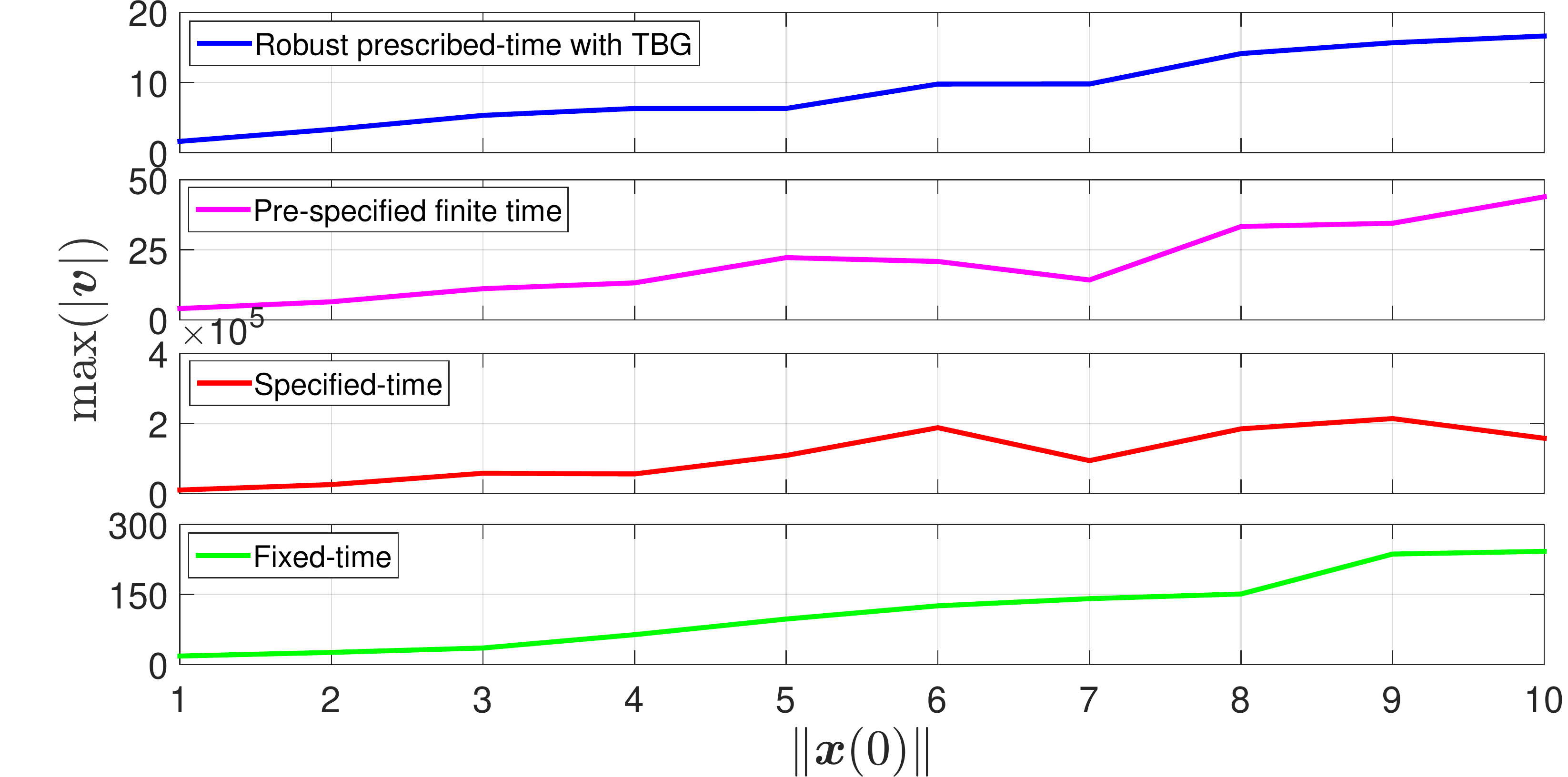}
\includegraphics[width=8.0cm]{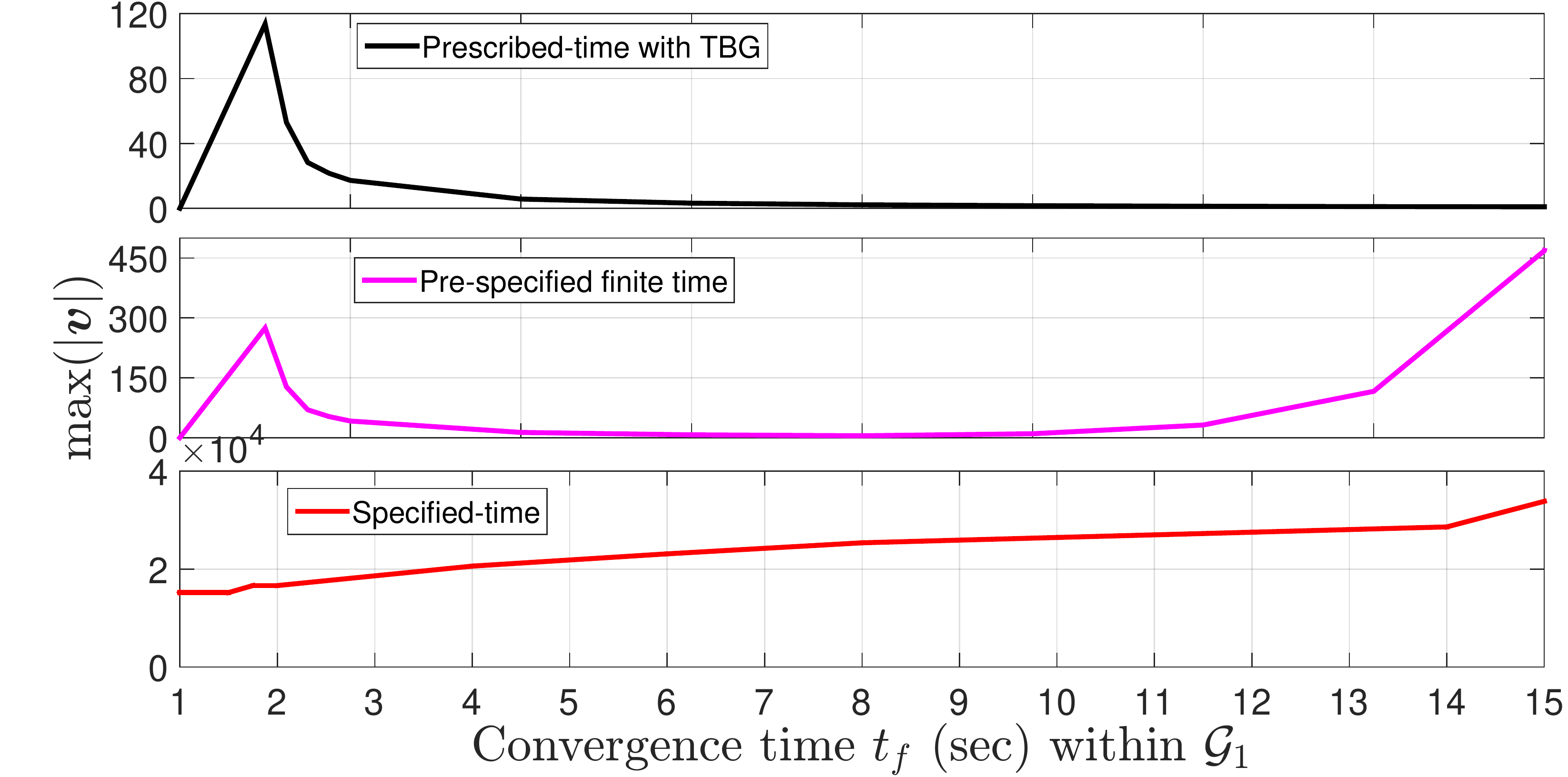}
\caption{
Comparison of the maximum value of the absolute auxiliary control input $\max(|\bld{v}|)$ as a function of the initial condition $\bld{x}(0)$ (top) and convergence time $t_f$ (bottom). The proposed TBG-based~\eqref{stc_control}, the Pre-specified finite time (\cite{Y.Wang_Y.Song}) and the Specified-time~(\cite{Y.Zhao_Y.Liu_G.Wein_W.Ren_G.Chen}) controllers are able to maintain the same convergence time for all the initial conditions. This is not the case for the Fixed-time~(\cite{Z.Zuo_B.Tian_M.Defoort_Z.Ding}) controller which has different increasing settling times in the range from $5.1$ to $5.8$ for the different tested initial conditions.
}
\label{fig_Comparison_Undigrapg}
\end{centering}
\end{figure}

In order to widely compare the performance of the proposed protocol and the existing ones~(\cite{Y.Wang_Y.Song,Y.Zhao_Y.Liu_G.Wein_W.Ren_G.Chen,Z.Zuo_B.Tian_M.Defoort_Z.Ding}), simulations were performed with ten different initial conditions that were randomly selected for the eight agents in such a way that the norm of $\bld{x}(0)$ is varied from $1$ to $10$. The leader state was kept constant $\bld{x}_l=[-1,0,0]^T$ for our approach \eqref{eq:controller_lin} and for the leader-following schemes in~\cite{Y.Wang_Y.Song} and~\cite{Z.Zuo_B.Tian_M.Defoort_Z.Ding}. The same previous control gains and preset settling time $t_f= 5$s were used for all the cases with the graph $\mathcal{G}_1$ for~\cite{Y.Wang_Y.Song} and~\cite{Y.Zhao_Y.Liu_G.Wein_W.Ren_G.Chen}, and the same followers were used for~\cite{Z.Zuo_B.Tian_M.Defoort_Z.Ding}. For every experiment, the norm of the consensus error $\bld{e}(t_f)$ and the maximum absolute value of the auxiliary control input $\bld{v}$ were registered. During the simulations, all the controllers achieved consensus at the prescribed-time, with errors lower than ($||\bld{e}^f(t_f)||<1\times10^{-4}$). The results for the magnitude of the auxiliary control $\bld{v}$ are shown in fig. \ref{fig_Comparison_Undigrapg} (top). Notice that the maximum auxiliary control efforts are significantly lower for the TBG-based proposed controller with respect to the compared approaches (\cite{Y.Wang_Y.Song,Y.Zhao_Y.Liu_G.Wein_W.Ren_G.Chen,Z.Zuo_B.Tian_M.Defoort_Z.Ding}). Our prescribed-time scheme was also compared by computing the maximum value of the control input $\max(|\bld{v}|)$ for different values of the convergence time parameter $t_f$. The controller of~\cite{Z.Zuo_B.Tian_M.Defoort_Z.Ding} is not included in this comparison since it is not able to achieve a constant preset convergence time and is not straightforward to set a desired bound of the settling time; re-tuning of several parameters is needed. Such evaluation was performed using the graph $\mathcal{G}_1$ and the initial states given in Table \ref{x0_high_order_table}. The results are presented in fig. \ref{fig_Comparison_Undigrapg} (bottom), where it can be seen that the maximum control effort is lower for our proposed controller than using the compared approaches. In a range, the maximum control effort is lower as the convergence time increases for our prescribed-time approach and the pre-specified approach (\cite{Y.Wang_Y.Song}), but the relation is far away from being proportional. Surprisingly, the control effort was higher as the convergence time increases for the specified-time approach (\cite{Y.Zhao_Y.Liu_G.Wein_W.Ren_G.Chen}) along all the evaluated range. Further analysis is required to formulate relations between the maximum control effort, the convergence time and the initial consensus error in the proposed methodology.

\begin{figure}[ht]
\begin{centering}
\includegraphics[width=8.0cm]{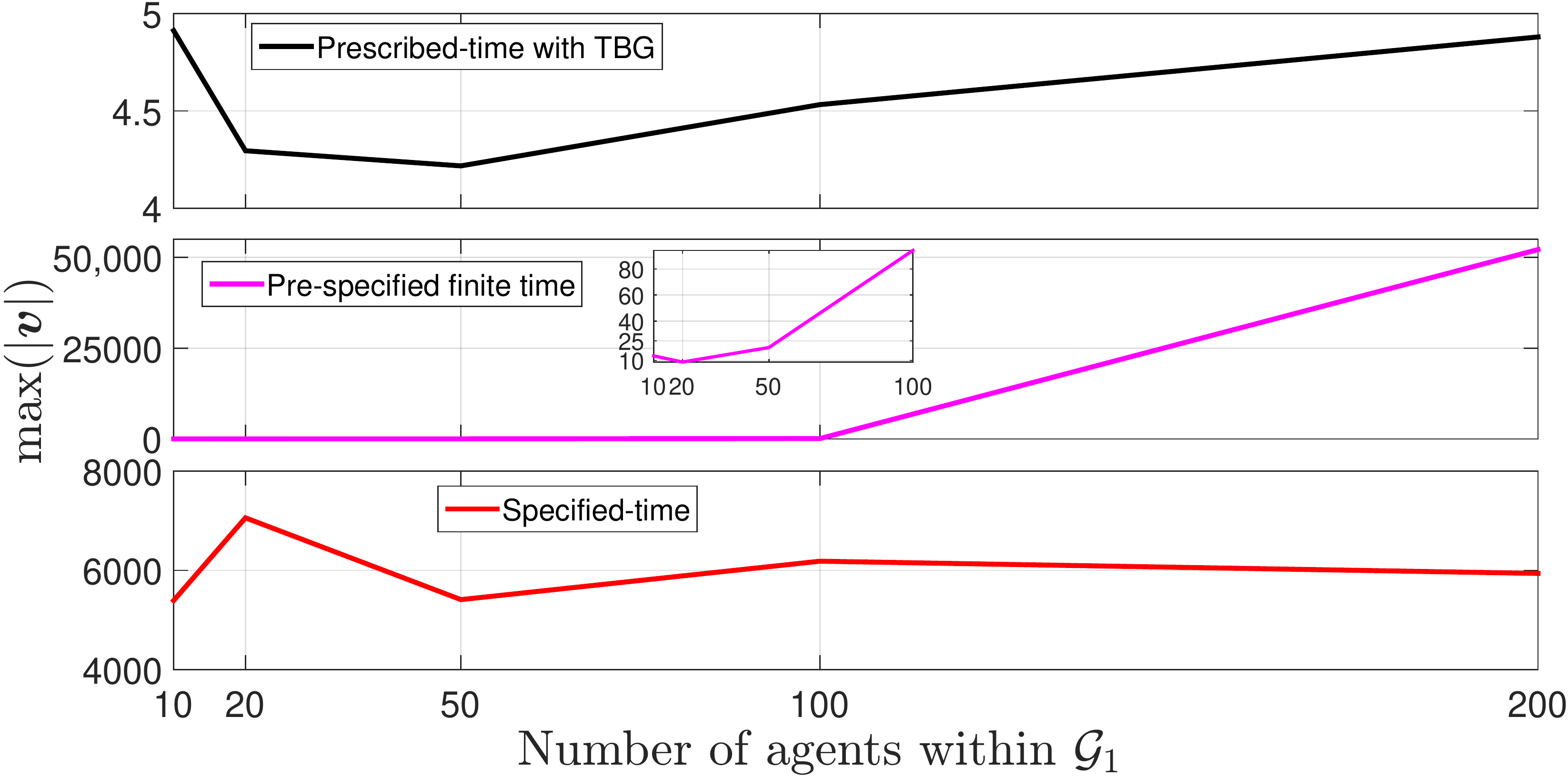}
\caption{Comparison of the maximum value of the absolute auxiliary control input $\max(|\bld{v}|)$ as a function of the number of agents. Controllers: Prescribed-time with TBG \eqref{eq:controller_lin}, Pre-specified finite time (\cite{Y.Wang_Y.Song}) and Specified-time (\cite{Y.Zhao_Y.Liu_G.Wein_W.Ren_G.Chen}).}
\label{fig_Comparison_NodesvsControl_Undigrapg}
\end{centering}
\end{figure}

Finally, another comparison was carried out with our approach and those of \cite{Y.Wang_Y.Song} and \cite{Y.Zhao_Y.Liu_G.Wein_W.Ren_G.Chen} by increasing the number of agents, in particular for cases with $10,20,50,100$ and $200$ second-order agents, considering a circular communication undirected graph (i.e., the $i$-th follower is connected to followers $i-1$ and $i+1$). The same initial state conditions were used for each approach, using values randomly selected between $-5$ to $5$. The second-order leader state was kept constant $\bld{x}_l=[-1,0]^T$ for our approach \eqref{eq:controller_lin} and the compared scheme of~\cite{Y.Wang_Y.Song}. The same previous control gains and preset settling time $t_f= 5$s were used for all the cases. It can be seen in fig. \ref{fig_Comparison_NodesvsControl_Undigrapg} that the control effort for our proposed controller~\eqref{eq:controller_lin} was almost constant for all the simulations and considerably lower than the compared controllers as the number of agents increases. Moreover, as the number of agents increases, the approach of \cite{Y.Zhao_Y.Liu_G.Wein_W.Ren_G.Chen} required readjustments of some control parameters to maintain the final consensus error lower than ($||\bld{e}^f(t_f)||<1\times10^{-2}$), which was not required for our proposed TBG controller.


\section{Conclusions}\label{sec:Conclusiones}

In this work, a couple of distributed control laws to achieve prescribed-time consensus have been proposed for a class of high-order MAS with nonlinear agents dynamics. 
The design of the proposed leader-following protocols combines the advantages of time base generators and feedback controllers to achieve closed-loop stability and robustness. The salient feature of this methodology is that consensus is achieved accurately in a prescribed time, independently of the initial conditions and detailed characteristics in the connectivity of the communication graph. Furthermore, the proposed leader-following protocol based on a sliding mode controller, provides robustness against matching disturbances, whilst the prescribed-time convergence property is maintained. Another advantage of the proposed method is that the generated control efforts can be continuous and smooth over time, exhibiting lower magnitudes than existing protocols for high-order prescribed-time (preset-time) consensus. Contrary to existing prescribed-time consensus protocols based on time-varying gains, in our approach no singularities appear in the control computation.

It is worth noting that the current results are valid for a MAS configuration with one leader and $N$ followers, assuming that the network topology contains a directed spanning tree, with the leader as the root. As a future direction of work, we are interested in extending the proposed methodology for MASs in a leader-less configuration and to the case where the MAS has more than one leader. In the last case, two problems can be addressed: the containment control where the followers must enter a target region formed by the group of leaders, and the problem of reaching multiple targets where the targets are specified by the state of multiple leaders and subgroups of followers must reach them. All these problems need further investigation to effectively extend our methodology to achieve the control objectives in a prescribed time. To deal with problems in the communication of the agents, we plan to consider switching topologies and time-delays in the MAS. Also, we are interested in addressing the case where the agents are time-varying systems.



\begin{IEEEbiography}[{\includegraphics[width=1in,height=1.25in,clip,keepaspectratio]{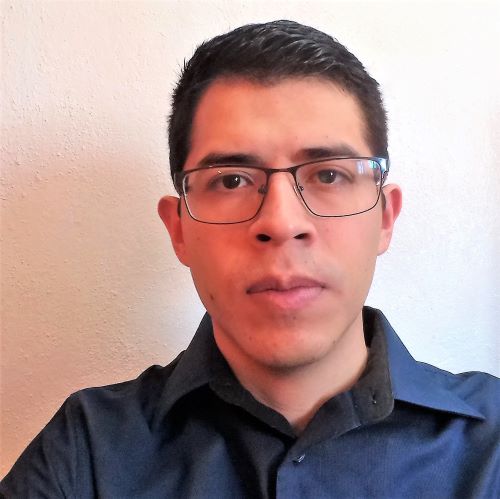}}]{J. ARMANDO COLUNGA} 
received the B.Sc. degree in mechatronics engineering from the Instituto Tecnol\'ogico y de Estudios Superiores de Monterrey (ITESM), campus Quer\'etaro, Mexico, in 2015 and the M.Sc. degree in computer science from the Centro de Investigaci\'on en Matem\'aticas, CIMAT-Guanajuato, Mexico, in 2018. 

He is currently a research associate with the Artificial Intelligence Consortium at CIMAT, Guanajuato, Mexico. His research interests include control theory and multi-robot systems, with particular focus on prescribed-time consensus algorithms in multi-agent systems.
\end{IEEEbiography}

\begin{IEEEbiography}[{\includegraphics[width=1in,height=1.25in,clip,keepaspectratio]{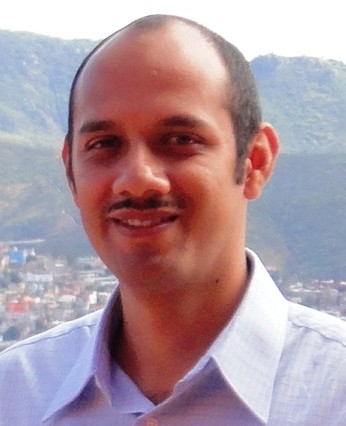}}]{H\'ector M. Becerra} (M'08) received a B.Sc. degree in electronics engineering from the Tecnológico Nacional de M\'exico, campus Ciudad Guzm\'an, a M.Sc. degree in Automatic Control from CINVESTAV-Guadalajara, Mexico, and a Ph.D. degree in systems engineering and computer science from the Universidad de Zaragoza, Spain, in 2003, 2005 and 2011, respectively. 

He is currently a full researcher at Centro de Investigación en Matemáticas, CIMAT-Guanajuato, Mexico. He is also member of the National System of Researchers, CONACyT, Mexico. His research interests include applications of control theory to robotics, particularly the use of computer vision as main sensory modality for feedback control of wheeled, humanoid and aerial robots, as well as control of multi-agent systems. 
\end{IEEEbiography}

\begin{IEEEbiography}[{\includegraphics[width=1in,height=1.25in,clip,keepaspectratio]{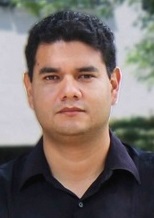}}]{Carlos R. V\'azquez} (M'20)
received the M.Sc. degree from CINVESTAV Unidad Guadalajara, Zapopan, Mexico, in 2006, and the Ph.D. degree in systems engineering from the Universidad de Zaragoza, Zaragoza, Spain. He is a Professor in Mechatronics at the Tecnologico de Monterrey Campus Guadalajara, Zapopan, Mexico. His research interests include the analysis and control of Petri nets and hybrid systems.
\end{IEEEbiography}

\begin{IEEEbiography}[{\includegraphics[width=1in,height=1.25in,clip,keepaspectratio]{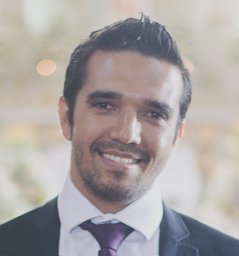}}]{David Gómez-Gutiérrez} (S'08-M'13-SM'17) received the M.Sc. and D.Sc. degrees in electrical engineering from CINVESTAV, Guadalajara, Mexico, in 2008 and 2013, respectively.

He is a Research Scientist at Intel Labs, Intel Tecnología de México and also an invited researcher in the Robotic's Focus Group at Tecnológico de Monterrey. He is currently a member of the National System of Researchers, CONACyT, Mexico.
His research interests include applications of control theory to robotics and multi-agent systems, as well as the development of predefined-time control algorithms.

\end{IEEEbiography}
\EOD

\end{document}